\documentclass[a4paper,11pt]{article}
\usepackage{amsmath,amssymb}
\usepackage{cleveref}
\usepackage{algorithm}
\usepackage{algorithmicx}
\usepackage{algpseudocode}
\usepackage{authblk}
\usepackage[
backend=biber,
style=numeric,
sortlocale=de_DE,
natbib=true,
url=false, 
doi=true,
eprint=false
]{biblatex}
\usepackage{bm}
\usepackage[final]{changes}
\usepackage{booktabs}
\usepackage{multirow}
\usepackage{caption}
\usepackage{subcaption}%for subfigures

\definechangesauthor[name=Jose,color=red]{JP}
\definechangesauthor[name=Philipp,color=blue]{PN}
\definechangesauthor[name=Markus,color=orange]{MM}

\newcommand{\rij}[2]{\operatorname{r}_{#1}^{#2}}
\newcommand{\coone}{\texttt{3c01}}
\newcommand{\coneeight}{\texttt{3c18}}
\newcommand{\czeroeight}{\texttt{3c08}}

\title{Linked Cell Traversal Algorithms for Three-Body Interactions in Molecular Dynamics}
\author[1]{Jos\'e Alfonso Pinz\'on Escobar}
\affil[1]{{Chair of High Performance Computing, Helmut Schmidt University, Germany}}
\author[2]{Markus Mühlhäu\ss er}
\affil[2]{Chair of Scientific Computing, Department of Computer Science, School of Computation, Information and Technology, Technical University of Munich}
\author[2]{Hans-Joachim Bungartz}
\author[3]{Philipp Neumann}
\affil[3]{{High Performance Computing \& Data Science, University of Hamburg, Germany
		   IT-Department, DESY, Germany}
		 }

\begin{document}
	\maketitle
	
	\begin{abstract}
	In this work, algorithms for the parallel computation of three-body interactions in molecular dynamics are developed.
	%This work proposes algorithms for the parallel computation of three-body interactions in molecular dynamics. 
	%The proposed linked cell traversals are based on already existing routines used for the computation of pair interactions.
	While traversals for the computation of pair interactions are readily available in the literature, here, such traversals are extended to allow for the computation between molecules stored across three cells.
	%Here, an extension to include the computation of forces between molecules distributed cross cell triplets is presented. 
	A general framework for the computation of three-body interactions in linked cells is described, and then used to implement the corresponding traversals.
	%In addition, our analysis is combined with the commonly used truncation conditions, which have been reported to have an impact on the total workload of the computations.
	In addition, our analysis is combined with the commonly used cutoff conditions, because they influence the total workload of the computation of interactions.
	\replaced[id=JP]{The combinations between traversals and truncation conditions are validated using the well-known Lennard-Jones fluid.
	Validation case studies are taken from the literature and configured into homogeneous and inhomogeneous scenarios.
	%Literature-based case studies are used in homogeneous and inhomogeneous scenarios. 
%	In homogeneous scenarios, ensemble averages are compared using literature-based case studies, and the corresponding radial distribution functions are provided as validation.
%	For the case of inhomogeneous molecules dynamics, density profiles are obtained using each traversal-truncation combination to show convergence in the generated interface geometries.
	}
	{
	The combination between traversals and truncation conditions are validated using the well-known Lennard-Jones fluid in single- and multi-phase case studies by comparing the obtained ensemble averages.	
	}
	Finally, strong scalability and performance in terms of molecule updates are measured at node-level.
	\par
	{\noindent\small \textbf{Keywords}:Molecular dynamics, three-body forces, linked cells, parallel computing}
	\par
%	\par
%	{\noindent\small \textbf{MSC codes}: 
%		35Q93, 49Q10, 65Y05, 65K10
%	}
	\par
	\end{abstract}
	
	\section{Introduction}\label{sec:intro}
	The calculation of intermolecular forces is a critical step in molecular dynamics simulations.
	%In molecular dynamics simulations, it is required to compute the forces between interacting particles. 
	These interactions are usually limited to pairs of molecules at short range, proof of which is the prevalent use of pair potentials such as Lennard-Jones, see e.g.~\cite{johnson1993,verlet1967,vrabec2006}.
	The main assumption being that higher order terms are negligible. 
	Nevertheless, studies have shown that for some cases this assumption is not entirely correct, for instance in the case of fluids in the liquid state~\cite{attard1992,sadus1998}.
	Therefore, in this work we provide algorithms for the parallel computation of non-additive three-body forces in molecular dynamics simulations limited to the case where the molecules are sorted using linked cells.
	%Additionally, we focus on exact calculations, i.e. without any estimations or approximations. 

	\noindent
	The existing literature focuses mainly on inhomogeneous scenarios, particularly on the vapor-liquid planar interface.
	One reason is related to the fact that three-body interactions, in the form of the Axilrod-Teller-Muto~\cite{axilrodteller1943,muto1943}, have been reported to contribute  between 5\% \-- 10\% of the pairwise potential energy in the liquid phase~\cite{barker1971,sadus1996}.  
	\added[id=JP]{As well as on the effect the Axilrod-Teller-Muto potential has on the computation of surface tension in simple liquids~\cite{barker1993}.
	Since it is reported that for substances like argon the error with respect to experimental results amounts to 19\% when using only two-body interactions, but reduces to about 2.2\% when three-body forces are accounted for.}
	
	Existing work can be found for the exact calculations of the three-body contributions to the phase behavior of argon~\cite{sadusPrausnitz1996}.
	Or the coexistence of the vapor-liquid phase of a Lennard-Jones fluid~\cite{sadus1998}.
	An approximation scheme is given in \cite{marcelliToddSadus2001,marcelliSadus2000}. where an empirical relation allows for the estimation of the three-body potential energy without requiring the explicit computations of the interactions at no cost in accuracy.
	
	From an algorithmic perspective, it is well-known that the computation of the forces accounts for most of the computational effort in a simulation. 
	The introduction of three-body interactions makes this even more noticeable. 
	In this sense, a common workload-reducing technique is the limitation of the range of the interactions by truncating them at a distance $r_c$, called the cutoff radius. 
	Varying approaches and distances for the three-body case can be found across the literature, see for instance~\cite{rittger1992,attard1992}.
	In recent work~\cite{nitzke2025}, these truncation conditions have been categorized into two major types, i.e. product and pair, described in \cref{sec:forceComp}. 
	However, the impact these conditions have on the total runtime and on the computed ensemble averages has not yet been addressed. 
	Another optimization related to the force computation, is the enforcement of Newton's 3rd law, $\vec{F}_{ij}=-\vec{F}_{ji}$, which reduces the computed forces by half. 
	In \cite{marcelliThesis}, the three-body forces are reformulated as force pairs to ease the implementation of Newton's 3rd law in the calculations.
	Other algorithmic approaches are based on the use of multiple time-stepping techniques in the context of parallel computing has been explored in \cite{martin2025,nakano1993} to reduce the computational effort associated to three-body interactions.
	As well as on strategies involving the use of algorithms for the parallel distribution of the force matrix~\cite{li2006,koanantakool2014}. 
	
	%End of introduction
	In this context, three-body interactions are important for the correct modeling of fluids in molecular simulations. 
	Given that the exact computation of these forces requires computationally intensive calculations, the optimization of this procedure is of high interest. 
	Therefore, in this work we provide an algorithmic description of how to perform these computations using parallel cell traversals in linked cells.
	These traversals are combined with the aforementioned optimizations, i.e. the truncation conditions and the enforcement of Newton's 3rd law. 
	%In what follows we provide an schematic description of the division of particle triplets in linked cells, and quantifying the total interactions that have to be computed for each case.
	In what follows, we describe the used potential, truncation conditions, and practical aspects of the computation.
	Then, we present a schematic description of how the three-body interactions of molecules stored in cells can be calculated and quantified.
	This is followed by the description of the cell traversals for the computation of intermolecular forces.
	Validation results using Lennard-Jones fluids are provided, together with a node-level performance analysis.
	Concluding remarks and a perspective for future studies complete this work.
	%Finally, concluding remarks and an outlook for future work wrap up our studies. 

		\section{Three-Body Computations}\label{sec:forceComp}

	\begin{figure}[!htbp]
		\centering
		\includegraphics[width=0.6\textwidth]{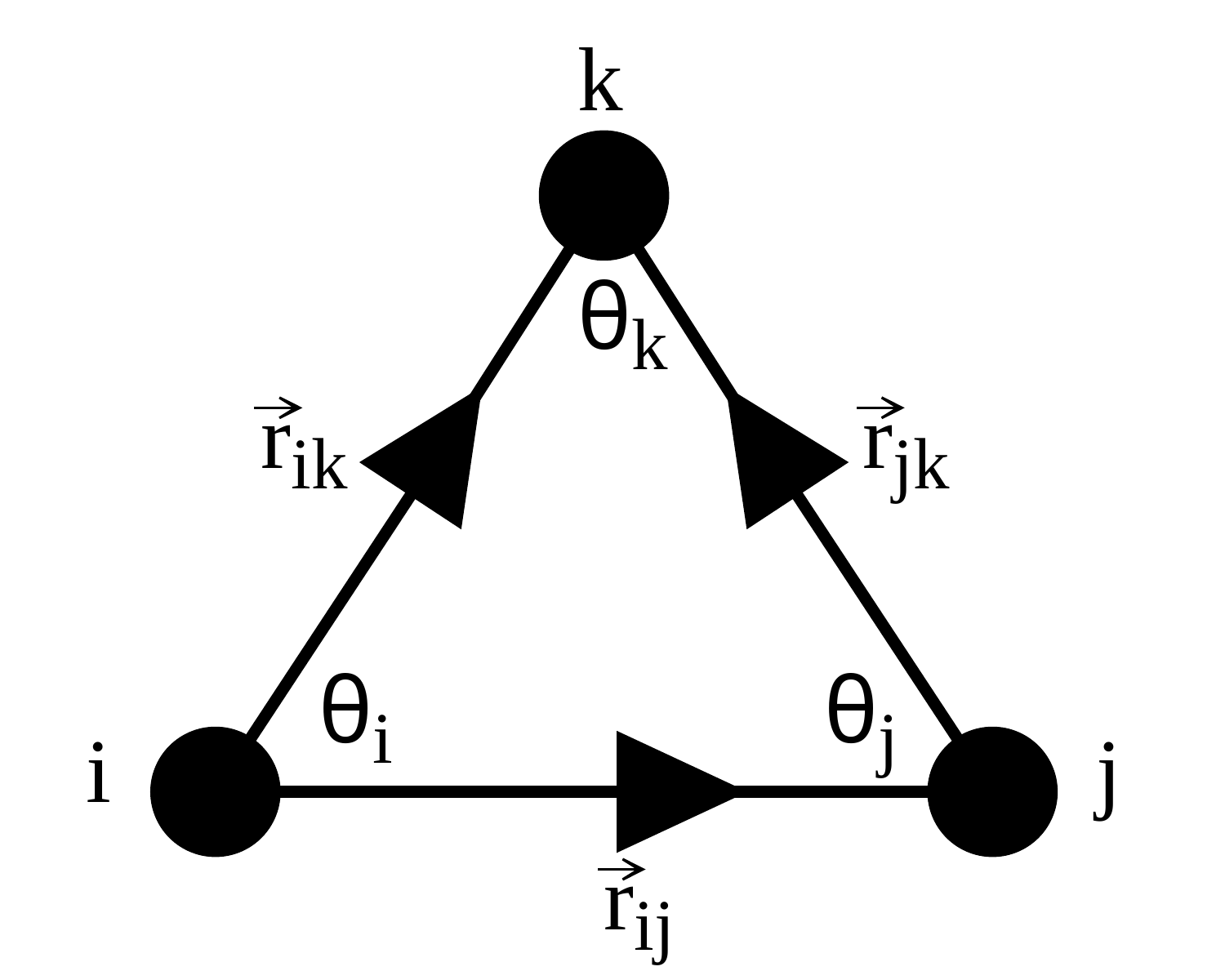}
		\caption{
			A molecular triplet is shown. 
			The distances follow the convention $\vec{r}_{\alpha\beta}=\vec{r}_\alpha-\vec{r}_\beta$. 
			The potential is defined in terms of the magnitudes of the relative distances.
		}
		\label{fig:triangle}
	\end{figure}
	
	In this section we describe practical aspects concerning the computation of three-body interactions.
	%A molecular triplet is shown in \cref{fig:triangle}, as before the indices of the three molecules $i,j,k$ are interchangeable.
	A molecular triplet $i,j,k$ is given in \cref{fig:triangle}, where the indices are interchangeable.
	Thus, the interaction depends on the distances between the particles, i.e. $\rij{ij}{},\rij{jk}{},\rij{ik}{}$.
	Using these distances and based on \cite{nitzke2025}, two major truncation conditions have been defined as:
	\begin{enumerate}
		\item \textbf{Pairwise cutoff}: Truncation if $\left(r_{ij}\vee r_{jk}\lor r_{ik}\right)>r_c$, such that the interactions are not computed if any of the sides of the triangle are larger than $r_c$. This has been suggested by \cite{allen&tildesley2017,attard1992} for the implementation of the minimum image convention with a cubic box and periodic boundary conditions. 
		\newline
		\item \textbf{Product cutoff}: Truncation if $\rij{ij}{}\rij{ik}{}\rij{jk}{}>r_c^3$, for which the truncation occurs if the product of the triangle sides is larger than the cube of the cutoff radius. This was suggested in \cite{rittger1992} as the three-body equivalent of $rr<r_c^2$, and was recently explored in the context of long-range corrections in \cite{nitzke2025}.
	\end{enumerate}
	
		\begin{figure}[!htbp]
		\centering
		\includegraphics[width=0.4\textwidth]{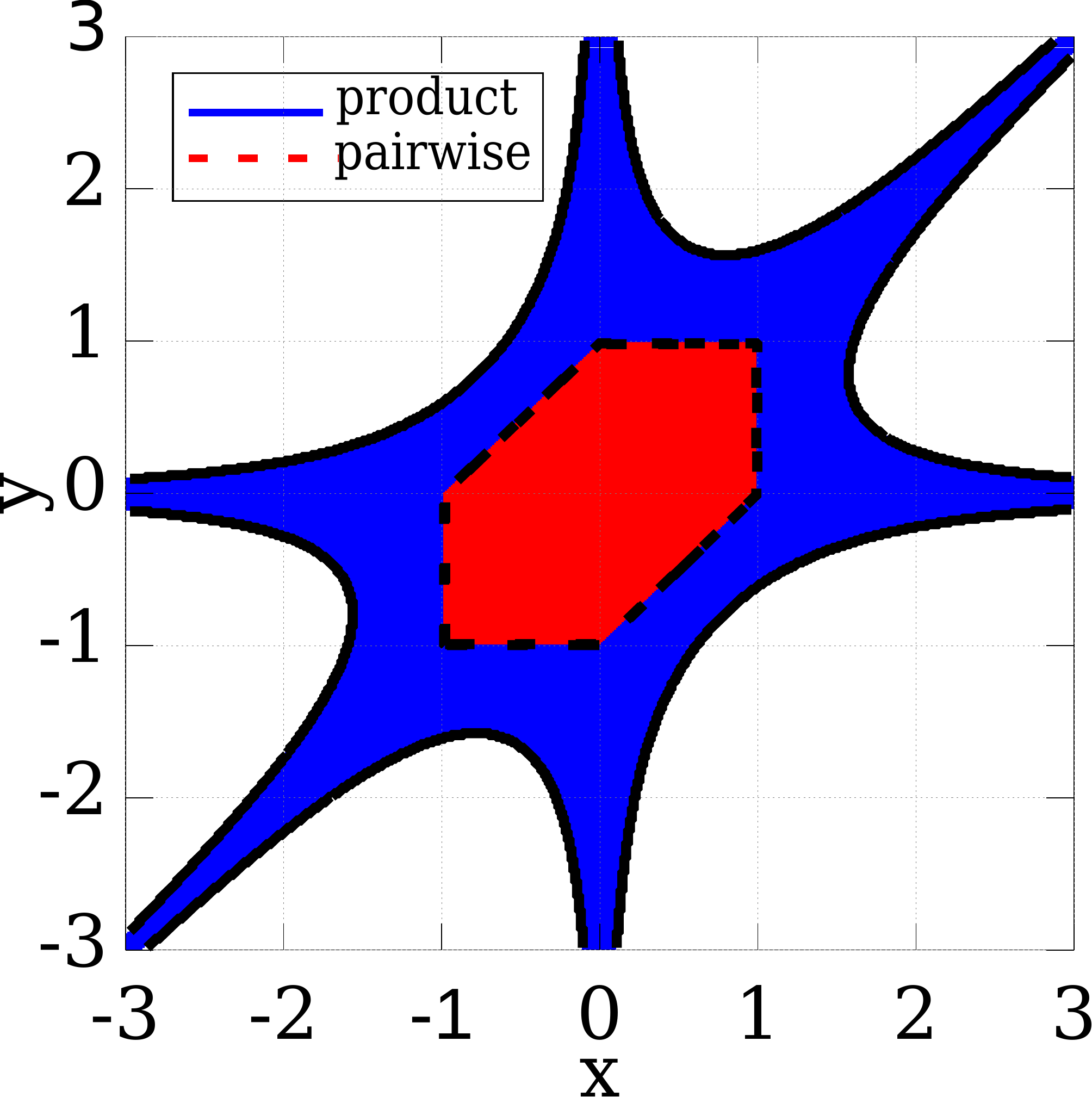}
		\caption{
			All valid triplets for the pairwise cutoff (red, inner area surrounded by segmented line) and product cutoff (blue, outer region surrounded by solid line) in the $xy$ plane for $r_c = 1$.
		}
		\label{fig:3body_surface}
	\end{figure}
	
	\replaced[id=MM]{
		Generally speaking, the pairwise cutoff is more restrictive than the product.
		As it is indicated in \cref{fig:3body_surface} for a cutoff of $r_c = 1$, the pair cutoff defines a subset of valid particles of the product cutoff.
	}
	{
		Generally speaking, the pairwise cutoff is more restrictive than the product which can be seen
	}
	All valid triplets for the pairwise cutoff are also valid for the product cutoff, but not vice versa.
	This means, a higher workload is expected when using the product cutoff.
	Thus truncations are combined with the traversals to determine their effect on both the performance and the ensemble averages.
	
	Based on the scheme of \cref{fig:triangle}, the forces on particle $i$ can be understood as the force acting from particle $j$ and another force from particle $k$.
	This logic can be extended to all three molecules.
	In \cite{marcelliThesis}, this has been done to express the forces of $i,j$, $k$ in pairs given as
	\begin{equation}
		\begin{aligned}\label{eq:forcepairs}
			\vec{F}_{i\left(jk\right)} &= \mathrm{\color{white}+}\vec{F}_{i\left(j\right)k} + \vec{F}_{i\left(k\right)j}\\
			\vec{F}_{j\left(ik\right)} &= -\vec{F}_{i\left(j\right)k} + \vec{F}_{j\left(k\right)i}\\
			\vec{F}_{k\left(ij\right)} &= -\vec{F}_{i\left(k\right)j} - \vec{F}_{j\left(k\right)i},
		\end{aligned}
	\end{equation}
	where, e.g., $F_{i\left(jk\right)}$ is the total force on particle $i$ due to $j$ and $k$, and $F_{i\left(j\right)k}$ the force on particle $i$ due to $j$.
	From this, Newton's 3rd law can be formulated for a molecule triplet as
	\begin{equation}
		\begin{aligned}\label{eq:newton3rdlaw}
			\vec{F}_{i\left(jk\right)} &+ \vec{F}_{j\left(ik\right)} &= -\vec{F}_{k\left(ij\right)}.
		\end{aligned}
	\end{equation}
	From \eqref{eq:forcepairs} and \eqref{eq:newton3rdlaw}, it is seen that only three forces are required to enforce Newton's third law.

	Let us define a potential, $u^{3\mathrm{body}}_{ijk}$, for the interaction between three molecules.
	Here we make use of the Axilrod-Teller-Muto potential as given by \cite{marcelliThesis}, where the dependence on the angles $\theta_{i},\theta_{j},\theta_{k}$ has been removed using the law of cosines
	\begin{equation}\label{eq:atm}
		\begin{split}
			u_{ijk}^{3\mathrm{body}}=\nu& \{\frac{1}{\rij{ij}{3}\rij{ik}{3}\rij{jk}{3}}\,\\
			&+\,3\frac{\left(-\rij{ij}{2}+\rij{ik}{2}+\rij{jk}{2}\right)\left(\rij{ij}{2}-\rij{ik}{2}+\rij{jk}{2}\right)\left(\rij{ij}{2}+\rij{ik}{2}-\rij{jk}{2}\right)}{8\,\, r_{ij}^5r_{ik}^5r_{jk}^5}
			\}.
		\end{split}
	\end{equation}
	Then the forces are defined as the gradient of \eqref{eq:atm} with respect to each of the distances associated to each particle.
	For instance, the forces on particle $i$ would require $\vec{F}_{i\left(jk\right)}=\vec{\nabla}_{\vec{r}_{ij}}u_{ijk}^{3\mathrm{body}}+\vec{\nabla}_{\vec{r}_{ik}}u_{ijk}^{3\mathrm{body}}$.
	Finally, the corresponding derivative to each necessary force can be computed by cycling the indeces $i,j,k$ from
	\begin{equation}
		\begin{split}\label{eq:ATMGradient}
			\frac{\partial u_{ijk}}{\partial r_{ij}} = 3\nu \{  &-\frac{1}{\rij{ij}{4}\rij{ik}{3}\rij{jk}{3}} -\frac{1}{8\rij{ik}{5}\rij{jk}{5}}+\frac{5\rij{ik}{}}{8\rij{ij}{6}\rij{jk}{5}}+\frac{5\rij{jk}{}}{8\rij{ij}{6}\rij{ik}{5}} \\
			&-\frac{1}{8\rij{ij}{2}\rij{ik}{3}\rij{jk}{5}} -\frac{1}{8\rij{ij}{2}\rij{ik}{5}\rij{jk}{3}} -\frac{3}{8\rij{ij}{4}\rij{ik}{}\rij{jk}{5}}\\ &-\frac{3}{8\rij{ij}{4}\rij{ik}{5}\rij{jk}{}} 
			-\frac{5}{8\rij{ij}{6}\rij{ik}{}\rij{jk}{3}}-\frac{5}{8\rij{ij}{6}\rij{ik}{3}\rij{jk}{}}+\frac{6}{8\rij{ij}{4}\rij{ik}{3}\rij{jk}{3}}\}.
		\end{split}
	\end{equation}
	Using \eqref{eq:ATMGradient}, the forces on the right-hand side of \eqref{eq:forcepairs} can be computed.

	\section{Particle Triplets in Linked Cells}\label{sec:generalFrame}
	
	In this section, the calculation of three-body interactions in linked cells is described using a general algorithmic framework, assuming the enforcement of Newton's third law throughout the calculations.
	In this way, only unique molecule triplets are accounted for.
	We propose a division of the computations into individual routines, based on the molecules being stored in one, two, or three different cells. 
	We limit the computed interactions to the immediate neighboring cells around cell $C_b$, called from here on the base cell. 
	As opposed to the case of only pair interactions, where this could be straightforwardly deduced, computing the forces between three-particles requires a deeper analysis.
	Furthermore, the different routines required for a correct computation of three-body interactions are illustrated here. 
	
	\begin{figure}[!htbp]
		\centering
		\includegraphics[width=0.15\textwidth]{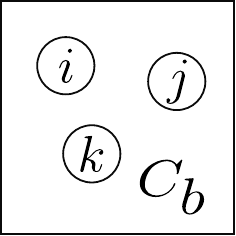}
			\caption{
			The base cell, $C_b$, contains three molecules $i,j,k$ such that $N_b=3$, for which the total interactions are given by \eqref{eq:celltriplet}. 
			}
		\label{fig:oneCell}
	\end{figure}
	
	%One cell
	\paragraph{One Cell}
	We start by computing the interactions within a single cell, $C_b$.
	%This is illustrated in \cref{fig:oneCell}, where $C_b$ is depicted with $N_b=3$ total number of molecules, which are $i,j,k$ for simplicity.
	This is illustrated in \cref{fig:oneCell}, where $C_b$ is depicted with a single molecule triplet $i,j,j$ such that the total number of molecules stored in $C_b$,  $N_b=3$, for simplicity.
	In this context, the total number of unique triplets in $C_b$ is given by
	\begin{equation}\label{eq:celltriplet}
		\left<C_b\right>:= \binom{N_b}{3}=\frac{N_b(N_b-1)(N_b-2)}{6}
	\end{equation}
	where the product $\left<..\right>$ is defined here as the total, unique three-body interactions for a given number of cells.
	
	\begin{figure}[!htbp]
		\centering
		\includegraphics[width=0.3\textwidth]{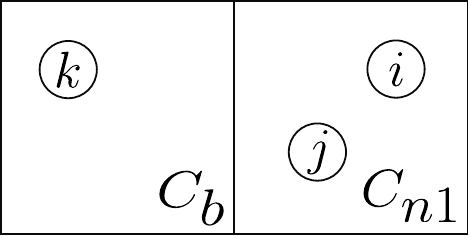}\hspace{0.5cm}
		\includegraphics[width=0.3\textwidth]{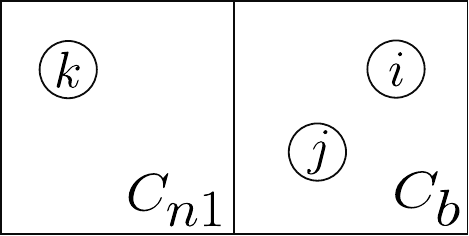}
		\caption{
			An atomic triplet is split between the base cell, $C_b$, and an immediate neighbor $C_{n1}$.
			It is first assumed that two molecules are stored in $C_b$, then this assumption is reversed to $C_{n1}$.
			%Two possible cases are shown, such that the cell pair is to be processed twice under two different assumptions of where two of the molecules are stored. 
		}
		\label{fig:2cells}
	\end{figure}
	\paragraph{Two Cells}
	Then we move on to the case where the three molecules are stored in two adjacent cells. 
	Assume that the triplet $i,j,k$ is split between $C_b$ and $C_{n1}$, the latter is one of the immediate neighbors of the former. 
	This is shown in \cref{fig:2cells}.
	First, account for the case when two particles are stored in $C_{n1}$ and one in $C_{b}$.
	After, the case where two molecules are stored in $C_b$.
	Let $N_{n1}$ be the number of molecules stored in $C_{n1}$, and the total unique triplets be given by
	\begin{equation}\label{eq:twoonetriplets}
		\left<C_b,C_{n1}\right> :=\, N_{n1}\frac{N_b(N_b-1)}{2}\,+N_b\,\frac{N_{n1}(N_{n1}-1)}{2},
	\end{equation}
	where as before $\left<C_b,C_{n1}\right>$ represents the total number of interactions between the $C_b$ and $C_{n1}$. 
	These cases also show that the same cell pair is to be used twice during the computations. 

	\begin{figure}[!htbp]
		\centering
		\includegraphics[width=0.3\textwidth]{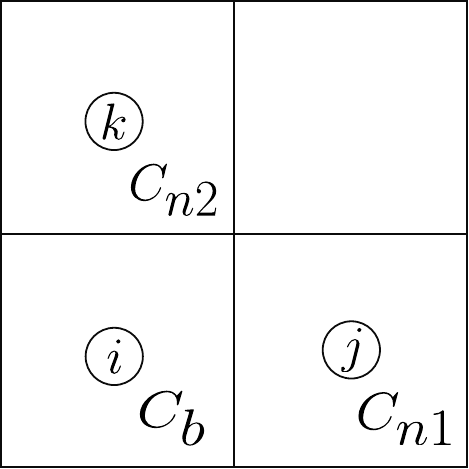}
			\caption{
			The particle triple $i,j,k$ is split between the cell triplet $\left(C_b,C_{n1},C_{n2}\right) $.
			A routine must be used to find a valid cell triplet using the base pair $\left(C_b,C_{n1}\right)$.
			}
		\label{fig:3cells}
	\end{figure}
	\paragraph{Three Cells}
	Finally, the case where each molecule is stored in a different cell is to be considered.
	Similar to the previous cases, in \cref{fig:3cells} this is shown with a molecular triplet $i,j,k$ being distributed across three different cells.
	As before, the computations are with respect to $C_b$, for which a first neighbor $C_{n1}$ has been found.
	Then, it would be required to find an additional and different neighboring cell, $C_{n2}$, to form a valid cell triplet with. 
	Assuming that $N_{n2}$ represents the total number of molecules in $C_{n2}$, the total interactions are given by the Cartesian product of the set of particles of each cell as
	\begin{equation}\label{eq:carttriplets}
		\left<C_b,C_{n1},C_{n2}\right> = N_b\cdot N_{n1}\cdot N_{n2}.
	\end{equation}
	\vspace{1cm}
	 
	The proposed framework allows for the partitioning into smaller routines of the total, unique three-body interactions between $N=N_b+N_{n1}+N_{n2}$ molecules as given by
	\begin{equation}\label{eq:totalInteractions}
		\begin{aligned}
			%\binom{N}{3}:=
			\frac{N\left(N-1\right)\left(N-2\right)}{6} =&\left<C_b\right>+\left<C_{n1}\right>+\left<C_{n2}\right>\\ &+\left<C_b,C_{n1}\right>+\left<C_b,C_{n2}\right>+\left<C_{n1},C_{n2}\right>\\
			&+\left<C_b,C_{n1},C_{n2}\right>.
		\end{aligned}
	\end{equation} 
	making use of \eqref{eq:celltriplet}, \eqref{eq:twoonetriplets}, and \eqref{eq:carttriplets} from above for all the possible cases between cells $C_b$, $c_{n1}$, and $C_{n2}$. 
	%Moreover, from \eqref{eq:totalInteractions} it is seen that the enforcement results in a 
	The process of finding all cell and molecular triplets is summarized in \cref{alg:generalScheme}.
	
	\begin{algorithm}
		\caption{Routines required for computation of three-body forces in linked cells}
		\label{alg:generalScheme}
		\begin{algorithmic}[1]
			\Require $N_c$ cells in linked cells
			\For{every cell $C_b\in N_c $}
			\State Compute base cell interactions $\left<C_b\right>$
			\State Find valid, immediate neighbor $C_{n1}$
			\State Compute cell pair interactions $\left<C_b,C_{n1}\right>$
			\State Retain base pair $\left(C_b,C_{n1}\right)$
			\State Find second valid neighbor $C_{n2}$ 
			\State Compute cell triplet interactions $\left<C_b,C_{n1},C_{n2}\right>$
			\EndFor
		\end{algorithmic}
	\end{algorithm}
	
%	Using these scheme, all the interactions between unique molecular triplets stored in three cells can be computed.
%	This is equivalent to saying that the total number of interactions for $N=N_b+N_{n1}+N_{n2}$ is given by
%	
%	
%	such that the given scheme splits the computation into different cases. 
%	On the other hand, \eqref{eq:totalInteractions} illustrates that for the case of no enforcement of Newton's third law, non-unique triplets are accounted for resulting in six-fold increase in the total interactions.
%	
	For a set of $N_c$ linked cells arranged in a regular grid, the three different computation cases have to be performed. Lines 3 and 4 imply that cell pair traversals can be used as a starting point for three-body computations.
	%Once a cell pair is found, it is used as a base for traversing all cell triplets.
	It is important to notice that the computation in line 5 retains the initial base pair, i.e. $\left(C_b,C_{n1}\right)$, while it traverses the remaining neighbors finding all the required cell triplets.
	Therefore, obtaining the neighbor $C_{n2}$ and computing the interactions, as in lines 6 and 7, requires appropriately designed routines.
	Fulfilling the equality in \eqref{eq:totalInteractions} depends on a correct traversal of the neighboring cells and an adequate definition of what a valid neighbor is.
	In what follows, it is explained that such a valid neighboring cell is related to traversing only forward and contiguous neighbors. 
	
%	
%	Finally, the workflow is summarized in \cref{alg:generalScheme}.
%	For a total set of $N_c$ linked cells, each of the three computations given by \eqref{eq:celltriplet},\eqref{eq:twoonetriplets}, and \eqref{eq:carttriplets} has to be performed.
%	In lines 3 and 4, a cell pair is required, this is already available using existing cell pair traversals.
%	Once the base pair has been found, line 5, a routine is required which allows for finding the neighbor $C_{n2}$ and forming the corresponding cell triplet. 
%	Therefore, \cref{alg:generalScheme} emphasizes that preexisting pair traversals can be reused and extended to account for three-body forces.
%	And that a routine to obtain the second neighbor, $C_{n2}$, while enforcing Newton's third law could potentially decrease the total number of traversed molecular triplets by a factor of 6. 
%	The design and implementation of these routines are described in what follows. 

	\section{Cell Traversals for Three-body Interactions}\label{sec:traversals}
	
	%Preamble
	The traversal routines allow for a parallel computation of the molecular interactions within linked cells,~\cite{tchipevC042019,tchipevC08Sliced2015,gratl2022}.
	Appropriate coloring patterns have to be used to avoid race conditions, by distributing the workload per color over several cores. 
	In this section we describe how, using the described framework, the required cell triplets are found.
	This is done using the cell pair traversals as a starting point since, as explained in \cref{sec:generalFrame}, finding cell triplets follows directly from two-body case. 
	A lexicographical ordering is assumed for all cells in the data structure and the interactions are limited to those between immediately neighboring cells.
	The latter implies that the cell size is equal to $r_c$, the cutoff radius.

	%Naive approach
	\paragraph{3c01 traversal} 
	\begin{figure}[!htbp]
		\centering
		\includegraphics[width=0.3\textwidth]{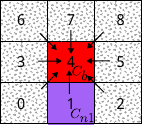}
		\hspace{0.5cm}
		\includegraphics[width=0.3\textwidth]{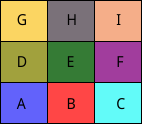}
		\caption{
			The \coone traversal and its coloring pattern. 
			All the triplets with respect to a base pair, here $(C_b,C_{n1}):=(4,1)$, would have to be found from the forward and backward neighbors.
			One color is assigned per cell, allowing for high parallelism, but without optimizing the computation of the forces by enforcing Newton's third law. 
		}
		\label{fig:c01traversal}
	\end{figure}
	An approach that does not account for Newton's third law but can be used as a reference is based  on traversing all neighbors of $C_b$ in a double loop, to find all possible cell triplets. 
	%A first, naive approach is based on a brute force method of traversing twice all the immediate neighbors of $C_b$, as to find all the possible cell triplets. 
	Such a traversal has been explored in \cite{martin2025}, where the valid cell triplets are limited to those within a certain cutoff distance, so that not all possible interactions are accounted for. 
	Our proposed routine is shown in \cref{fig:c01traversal}, where the base pair $\left(C_b,C_{n1}\right):=\left(4,1\right)$, is marked together with the neighborhood to be traversed. 
	In this case, all the triplets would have to be found as $C_{n2}:=\left\{0,2,3,5,6,7,8\right\}$, i.e. by traversing all the forward and backward neighbors.
	Because Newton's 3rd law is not enforced, all cell triplets are traversed more than once.
	Only the forces from the molecules in $C_b$ are computed, which leads to finding each molecular triplet repeatedly and performing redundant calculations. 
	One advantage, as seen in \cref{fig:c01traversal}, is that all the interactions can be calculated simultaneously, by assigning one color to every cell. 
	
	The lack of enforcement of Newton's third law hampers down the achievable performance, therefore more elaborate traversals are required.
	An alternative is then to traverse only the set of unique cell triplets around $C_b$, as described in \cref{sec:generalFrame}.
	As in the cell pair case, reducing the traversed neighbors with respect to $C_b$, is one way to achieve a higher performance. 
	This is done by traversing only forward neighbors, which in lexicographical terms implies $c_1<c_2$ with $c_1$ and $c_2$ the indices of two neighboring cells. 
	Two routines of this type are detailed below.

\begin{figure}[!htbp]
		\centering
		\includegraphics[width=0.3\textwidth]{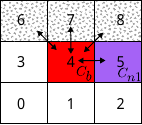}
		\includegraphics[width=0.3\textwidth]{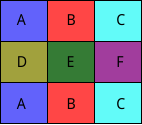}
		\caption{
			In order to implement Newton's 3rd law in the three-body force computation, the neighborhood is limited to only the forward neighbors of $C_b$. 
			For the base pair, $(4,5)$, finding the second neighbor $C_{n2}$ implies traversing only forward cells with respect to $C_{n1}$. 
		}
		\label{fig:c18traversal}
	\end{figure}

	\paragraph{3c18 traversal} 
	This routine traverses only the forward neighbors of $C_b$.
	This is shown in \cref{fig:c18traversal}, together with the coloring pattern that shows 6 and 18 colors are required, respectively, in 2d and 3d \cite{tchipevC08Sliced2015}. 
	Here, Newton's 3rd law is enforced -signaled by the bi-directional arrows in \cref{fig:c18traversal}.
	Once a cell pair is found, e.g. $\left(4,5\right)$, all the forward neighbors of $C_{n1}$ are traversed to form the cell triplets.
	Therefore, $C_{n2}:=\left\{6,7,8\right\}$, which implies a significant reduction of the traversed neighborhood.

\begin{figure}[!htbp]
		\centering
		\includegraphics[width=0.3\textwidth]{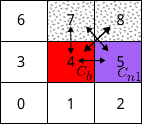}
		\includegraphics[width=0.3\textwidth]{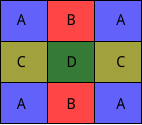}\\
		\vspace{0.1cm}
		\includegraphics[width=0.3\textwidth]{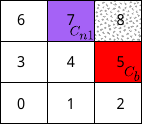}
		\includegraphics[width=0.3\textwidth]{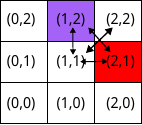}
		\caption{
			The neighborhood and triggered interactions are shown for the \czeroeight\, traversal, together with the coloring pattern in 2d, and the change of base cell.
			As an example, the new base cell $5$ is defined.
			First, a valid neighbor is used to define a base pair, here $(5,7)$. 
			Then, all the cell triplets are found with respect to it. 
			The local indices can be used to define a valid $C_{n1}$ of $C_b=5$ as those requiring a negative step in any principal direction. 
		}
		\label{fig:c08traversal}
	\end{figure}

	\paragraph{3c08 traversal} 
	Further reduction of the traversal region allows for a higher performance.
	The \czeroeight\, traversal is shown in \cref{fig:c08traversal}, where it is seen that $C_b$ interacts with fewer of its immediate neighbors.
	A more compact traversal allows for a higher workload per color~\cite{tchipevC08Sliced2015}, therefore this routine requires 4 and 8 colors in 2d and 3d, respectively. 
	Together with the enforcement of Newton's third law, a higher performance can be achieved at the cost of omitting certain interactions. 
	
	%	
	%	Another optimization can be obtained by a further reduction of the neighborhood, see also~\cite{tchipevC08Sliced2015}.
	%	As shown in \cref{fig:c08traversal}, the \czeroeight traversal reduces the neighborhood with respect to the base cell, $C_b$.
	%	Similar to the \coneeight traversal, the neighborhood from where $C_{n2}$ is found from reduced as the base pair move forward.
	%	It is seen that for the example base pair $(6,7)$, $C_{n2}:=\left\{9,10\right\}$ only.
	%	Such that a smaller neighborhood is used, for which 4 colors are required in 2d and 8 in 3d.

	In principle, this routine follows the same procedure as \coneeight, the forward neighbors of $C_b$ are used to define the base pairs, $\left(C_b,C_{n1}\right)$, and $C_{n2}$ is obtained among the forward neighbors of $C_{n1}$.
	In \cref{fig:c08traversal} this is shown as the base pair $\left(4,5\right)$ being used to compute the cell triplets with $C_{n2}:=\left\{7,8\right\}$ such that the traversed region is smaller. 
	One major difference is that \czeroeight\, allows for a change of $C_b$, see \cref{fig:c08traversal}.
	Once the new $C_b$ is determined, all valid base pairs are traversed to find unique cell triplets.
	%Using the new base cell, the cell triplets composed of $C_b$ and a backwards $C_{n1}$ have to be computed.
	%A backwards neighbor can be defined as any $C_{n1}$ which requires a negative step in terms of the local indices.
	A valid base pair is then defined as one where the corresponding $C_{n1}$ implies a negative step in terms of the local indices, see \cref{fig:c08traversal}.
	In this sense  for $C_b$ and $C_{n1}$ respectively, $\left(2,1\right)-\left(1,2\right)=\left(1,-1\right)$, would require a step in a negative principal direction. 
	Having found such a base pair, $\left(C_b,C_{n1}\right)$, $C_{n2}$ is found as before among the forward neighbors of $C_{n1}$. \\
	
	While the described traversals find all valid particle triplets according to the pair cutoff rule, it is important to note that not all triplets as defined by the product rule will be traversed.\comment[id=JP]{Maybe here refer to figure 2, since it illustrates this fact very clearly.}
	The product condition allows for triplets with side-lengths larger than $r_c$\added[id=MM]{, as it is seen in \cref{fig:3body_surface} for a cutoff of $r_c = 1$}.
	Therefore, interacting particles might be multiple cells apart. 
	The number of triplets found for the product rule is the highest for \coone, followed by \coneeight, and finally \czeroeight.

	% One disadvantage of this traversal is that by construction certain cell triplets are not traversed, as compared to \coneeight.
	% In \cref{fig:c08traversal} it can be seen that the cell triplet $\left(4,5,6\right)$ is not visited.
	% Therefore, a trade-off between performance and accuracy seems to be involved.
	% However, the use of the cutoff conditions given in \cref{sec:forceComp} might imply that these discarded interactions might not be critical to the estimation of ensemble averages
	% This is further explored in the following section.

	\begin{table}[!htbp]
		\begin{center}
		\begin{tabular}{cc|cc}
			\multirow{2}{*}{Traversal} & \multirow{2}{*}{Cutoff} & \multicolumn{2}{c}{Hitrate in \%} \\
			& & 3-body & 2-body \\
			\hline
			\multirow{2}{*}{\coone}
			&    Pair &1.19 & \multirow{2}{*}{15.52} \\
			& Product &7.50 \\
			\hline
			\multirow{2}{*}{\coneeight} 
			& Pair    &2.71 & \multirow{2}{*}{18.21} \\
			& Product &13.86 \\
			\hline
			\multirow{2}{*}{\czeroeight} 
			& Pair &4.00 & \multirow{2}{*}{18.21} \\
			& Product &20.34 \\
		\end{tabular}
		\caption{Hitrate for all traversals and cutoff conditions.
			The hitrate is defined as the ratio between the number of traversed triplets and those that actually contribute to the forces, i.e. fulfill the cutoff condition. The hitrate for corresponding pairwise traversals is shown as a comparison.
		}
		\label{tab:hitrate}
		\end{center}
	\end{table}

	\replaced[id=MM]{
			Another important aspect to consider is the ratio between the number of molecule triplets that eventually contribute to the forces, i.e. fulfill the cutoff condition, to the total number of traversed molecule triplets.
	}
	{
		Another important aspect to consider is the ratio between the number of traversed molecule triplets and those that actually contribute to the forces, i.e. fulfill the cutoff condition. 
	}
	We define this ratio as the hitrate of a traversal. 
	In \cref{tab:hitrate}, the hitrate is shown for all traversals and cutoff conditions.
	The numbers were obtained from simulations with uniformly distributed particles.
	It is seen that the hitrate increases when going from \coone\, to \coneeight\, and finally to \czeroeight\, due to the reduction of the traversed cells. 
	It is also seen that the product cutoff results in a much higher hitrate than the pairwise truncation as more triplets fulfill the cutoff condition.
	We will explore the effects of these differences in what follows.
	
	\section{Validation and Performance}\label{sec:results}
	In this section we present simulation results for the validation and performance measurement of the given traversals in \cref{sec:traversals} in combination with the cutoff conditions described in \cref{sec:forceComp}. 
	A single-site Lennard-Jones fluid is used in all cases, since its properties are well understood~\cite{johnson1993,verlet1967}.
	The pair interactions are given by Lennard-Jones potential, while \eqref{eq:atm} describes the three-body potential energy.
	\replaced[id=JP]{
	As previously mentioned, the introduction of three-body forces on the calculation of surface tension in vapor-liquid scenarios can reduce the overall error to about 2.2\% with respect to experimental data \cite{barker1993}.
	}
	{
	As previously mentioned, the contributions of the Axilrod-Teller-Muto potential to the properties of the phase space vary between vapor and liquid states.
	}
	For this reason, our validation results are based on homogeneous and inhomogeneous simulations.
	The latter are configured to form planar vapor-liquid interfaces~\cite{vrabec2006,holcomb1993,trokhymchuk1999}.

	Here on, all physical quantities are expressed using Lennard-Jones reduced units, assuming a fluid with $\epsilon=1,\sigma=1$, and a mass of $m=1$. 
	As in \cite[]{allen&tildesley2017}, the reduced quantities are given by  temperature $T^*=\kappa_B T/\epsilon$, density $\rho^*=\rho\sigma^3$, pressure $P^*=P\sigma^3/\epsilon$, potential energy $E^*=E/\sigma$, and time $t=(\epsilon/m)^{1/2}t$.
	For ease of notation, the superscript $^*$ is omitted from now on.
	Additionally, the non-additive parameter of the Axilrod-Teller-Muto potential is given as $\nu^*=\nu/\epsilon\sigma^9$ and set to $\nu^*=0.072$, as in \cite{attard1992}, for the remaining of this work. 
	
	The validation of the homogeneous results is based on measuring the potential energy per particle and pressure, $E/\mathrm{N}$ and $P$ respectively. 
	These are evaluated using
	\begin{equation}
		\begin{aligned}
			E\mathrm{/N} = &\left<E_2\right>\mathrm{/N}+\left<E_3\right>\mathrm{/N}\\
			P = &N\kappa_B T + \left<P_2\right> + \left<P_3\right>,
		\end{aligned}
	\end{equation}
	where $\left<E_2\right>$,$\left<P_2\right>$,$\left<E_3\right>$, and $\left<P_3\right>$ are the energy and pressure contributions of the pair and three-body interactions respectively. 
	Long-range corrections~\cite{allen&tildesley2017} are used in the homogeneous scenarios for the two-body ensemble averages, $\left<E_2\right>$ and $\left<P_2\right>$, while none are used for the inhomogeneous cases and in any of the three-body averages. 
	Furthermore, $P_3$ is evaluated with
	\begin{equation}\label{eq:P_3}
		\begin{aligned}
			P_3\,V=&\left<W_3\right>\\
			W_3=& \sum_{i}\sum_{j>i}\sum_{k>j}\left\{\vec{r}_{ij}\cdot\vec{F}_{i\left(j\right)k}+\vec{r}_{ik}\cdot\vec{F}_{i\left(k\right)j}+\vec{r}_{jk}\cdot\vec{F}_{j\left(k\right)i}\right\},
		\end{aligned}
	\end{equation}
	where $V$ is the volume of the domain and $W_3$ is the virial associated to the three-body interactions.
	Moreover, the computation of $W_3$ makes use of \eqref{eq:forcepairs}. 
	
	In the single-phase case studies, the results are compared to a pure Lennard-Jones simulation, i.e. $\nu=0$.
	\added[id=JP]{
	Additionally, we compare our ensemble averages to those given by \cite{attard1992}, for the used states.
	As a qualitative measure, the radial distribution functions for all cases are presented, and shown to converge to the same configuration. 
	Although it is expected for the three-body interactions to have some effect on the pairwise distribution functions, we did not observe major differences in the considered cases. 
	}
	%Whereas in the multi-phase scenarios the contributions from the three-body interactions are given with respect to the total pressure and energy per particle, $P$ and $E$/N, respectively.
	All ensemble averages are measured using the product and pairwise conditions for each traversal, in order to showcase their impact on the thermodynamic properties.
	\added[id=JP]{
	Whereas in the multi-phase simulations, results for the geometry of the planar interfaces are given via analyzing the obtained density profiles with the traversal-cutoff combinations.
	These are given together with the profile using only two-body interactions, i.e. without Axilrod-Teller-Muto forces.
	}
	Finally, the performance measurements for each combination of traversals and truncation conditions allow for a good understanding of the tradeoff between the effects on the ensemble averages and the workload optimization to reduce the runtime of the three-body calculations.
	These results were obtained using our in-house molecular dynamics code \texttt{LauraMD}~\cite{lauramd2025}, programmed in C++ with OpenMP parallelization.
	\subsection{Single-phase LJ Fluid}\label{ss:1phase}
	
	Meaningful thermodynamic states were tested for a pure LJ fluid in a cubic domain using periodic boundary conditions.
	These states are based on results provided by \cite{attard1992}.
	The simulations were set up inside a box domain of dimensions $a=12.5$.
	A cutoff radius of $r_c=2.5\sigma$ is used with a step size of $\Delta t=0.004$.
	Initially, the phase space is generated in a simple cubic lattice structure. 
	At first, only Lennard-Jones interactions are used for a melting phase of $25,000$ steps, followed by a $25,000$ production run from where the thermodynamic properties and radial distribution function are measured.
	The latter were sampled on every time step with a total of 600 bins. 
	These results are used as a reference, $\nu=0$, in \cref{tab:1phaseRes}.
	%An equilibration phase of $25,000$ steps is performed to bring the phase space to its liquid state using only two-body interactions.
	%Then, the reference Lennard-Jones properties and radial distribution function with $\nu=0$ were averaged over an additional $25,000$ steps.
	Finally, three-body forces are introduced to the calculations and the average total pressure and potential energy per particle are measured for an additional $25,000$ simulation steps, \added[id=JP]{together with the corresponding radial distribution function including the effects of three-body interactions.
	}

	\begin{table}[!htbp]
		\begin{center}
			\begin{tabular}{lc|cc|cc}
				Traversal & Cutoff & \multicolumn{4}{c}{State} \\
				&		   & \multicolumn{2}{|c}{T=1.033}& \multicolumn{2}{c}{T=0.746}\\
				&		   & \multicolumn{2}{|c}{$\rho=0.65$}& \multicolumn{2}{c}{$\rho=0.817$}\\
				&		   & \multicolumn{2}{|c}{N=1270}& \multicolumn{2}{c}{N=1596}\\
				\hline
				& & $E/\mathrm{N}$ & $P$ & $E/\mathrm{N}$ & $P$\\
				\multirow{2}{*}{\coone}
				&    Pair &-4.37965 & 0.10406 & -5.62938 & 0.48192 \\
				& Product &-4.37695 & 0.10933 & -5.62655 & 0.48764 \\
				\hline
				\multirow{2}{*}{\coneeight} 
				& Pair    &-4.38009 & 0.10682 & -5.63040 & 0.47474 \\
				& Product &-4.37872 & 0.11725 & -5.62686 & 0.48936 \\
				\hline
				\multirow{2}{*}{\czeroeight} 
				& Pair &-4.38044 & 0.10891 & -5.63252 & 0.47269 \\
				& Product &-4.37934 & 0.10601 & -5.63139 & 0.47222 \\
				\hline
				\multicolumn{2}{c|}{Attard\quad\cite{attard1992}}&-4.36&0.10&-5.64&0.38\\
				& & & & &\\
				& & \multicolumn{4}{c}{\underline{Lennard-Jones ($\nu=0)$}}\\
				& & -4.5395 & -0.1140 & -5.8932 & -0.09705 \\
				& & & & &\\
				& & \multicolumn{4}{c}{\underline{Lennard-Jones ($\nu=0)$ from Attard \cite{attard1992}}}\\
				%\multicolumn{2}{c|}{Attard\quad\cite{attard1992}}&  -4.52  & -0.11   & -5.90   & -0.20\\
				&&  -4.52  & -0.11   & -5.90   & -0.20\\
			\end{tabular}
			\caption{Validation results for a single-phase LJ fluid. 
				Different thermodynamic states in the NVT ensemble are tested for all traversals and truncations.
				The averages under the influence of three-body interactions are compared to two-body, $\nu=0$, results.
				The corresponding results reported in \cite{attard1992} are given as a reference.
			}
			\label{tab:1phaseRes}
		\end{center}
	\end{table}
	
	The results per traversal and cutoff condition are given in \cref{tab:1phaseRes}.
	%For these tests, a varying number of particles, $N$, has been used.
	As mentioned before, the obtained three-body and $\nu=0$ results are compared to those reported by \cite{attard1992}.
	%As mentioned before, the Lennard-Jones, $\nu=0$, and Attard\,\cite{attard1992} ensemble averages are given as a reference. 
	It is seen that the three-body contributions to the energy per particle result in more positive potential energy per particle.
	The pressure in the three-body cases is substantially higher than for the $\nu=0$ simulations.
	\added[id=PN]{Comparing our results to those reported in \cite{attard1992}, there seems to be strong consistency between them.
	However, some minor discrepancies with \cite{attard1992} are observed for the Lennard-Jones $P$ at $T=0.746$.
	These could be motivated by the differences in setup of the simulations, on the differences in long-range corrections used for $P$, as well as on numerical aspects like used $r_c$ and sampling length.
	On the other hand, the three-body contribution to $p$, $P_3$, is consistent with what is reported by \cite{attard1992}, thus validating our obtained ensemble averages.
	}
	\added[id=JP]{Good agreement between the averages reported in \cite{attard1992} and those provided in \cref{tab:1phaseRes} has been observed.
	}
	
	Some differences between the traversals for the same state are observed.
	The higher density $\rho=0.817$ shows some minor disagreements in the measured pressure, particularly for the \czeroeight\, case compared to \coneeight. 
	This could be related to the absence of certain cell triplets in the calculations, see \cref{sec:traversals}. 
	On the other hand, for the $\rho=0.65$ case there is strong agreement between all the measured values of $P$ and $E/\mathrm{N}$, with considerably consistent results even for the product- and pairwise-configured \coneeight\, scenarios. 
	The cutoff conditions seem to generate negligible differences between the different runs.
	Nevertheless, it is worth noting that most of the observed ensemble average variations, could be explained by the naturally occurring fluctuations proper of molecular dynamics simulations.
	So it is seen that consistent results were obtained overall with the expected differences proper of interaction truncations caused by the cutoff conditions and \coneeight\, traversal. 
	%Finally, here we observe that the additional molecular triplets introduced by the product cutoff, do not seem to cause dramatically different results in these single-phase case studies. 

	\begin{figure}
		\begin{center}
			\begin{tabular}{cc}
				Pairwise & Product \\
				\includegraphics[width=0.49\textwidth]{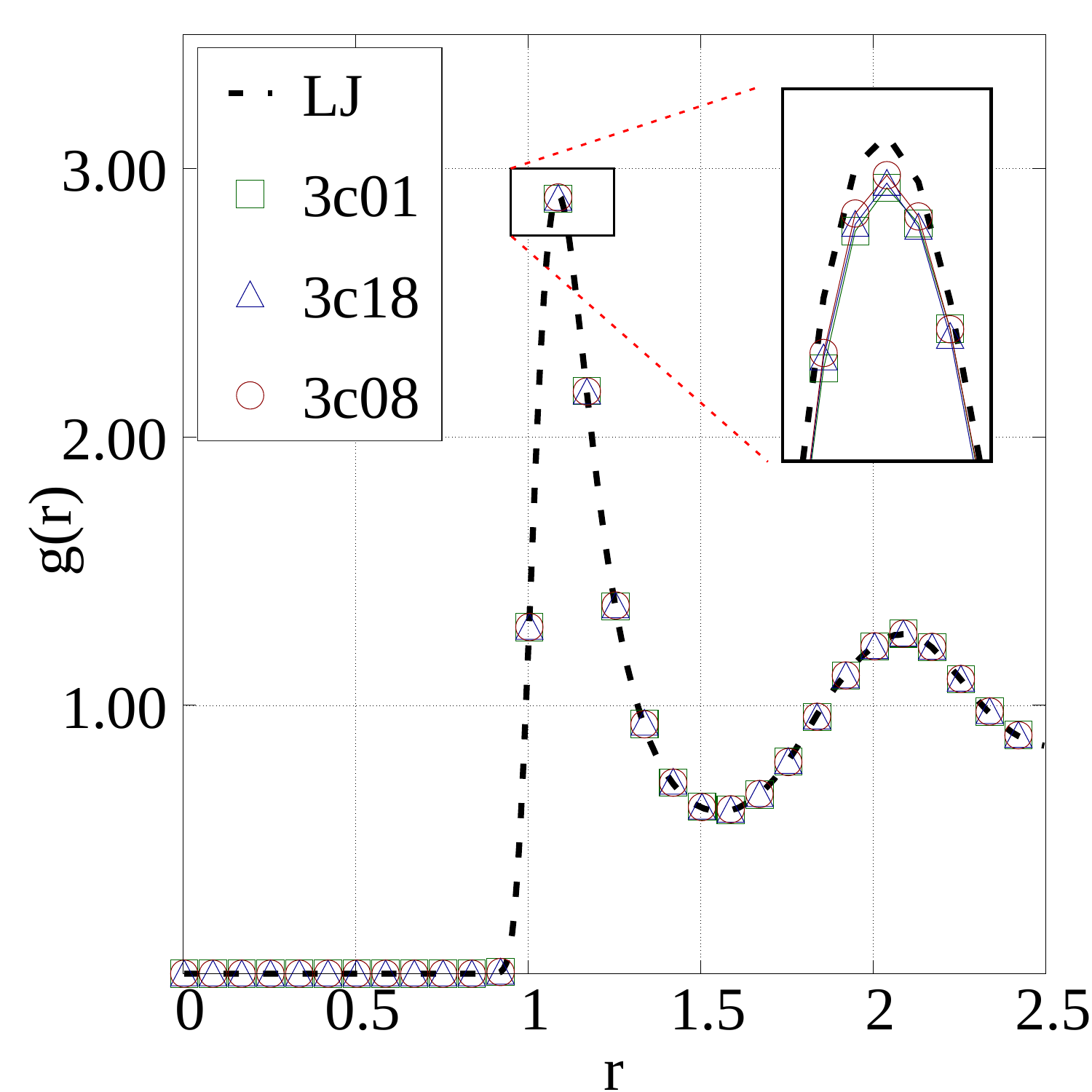}
				&
				\includegraphics[width=0.49\textwidth]{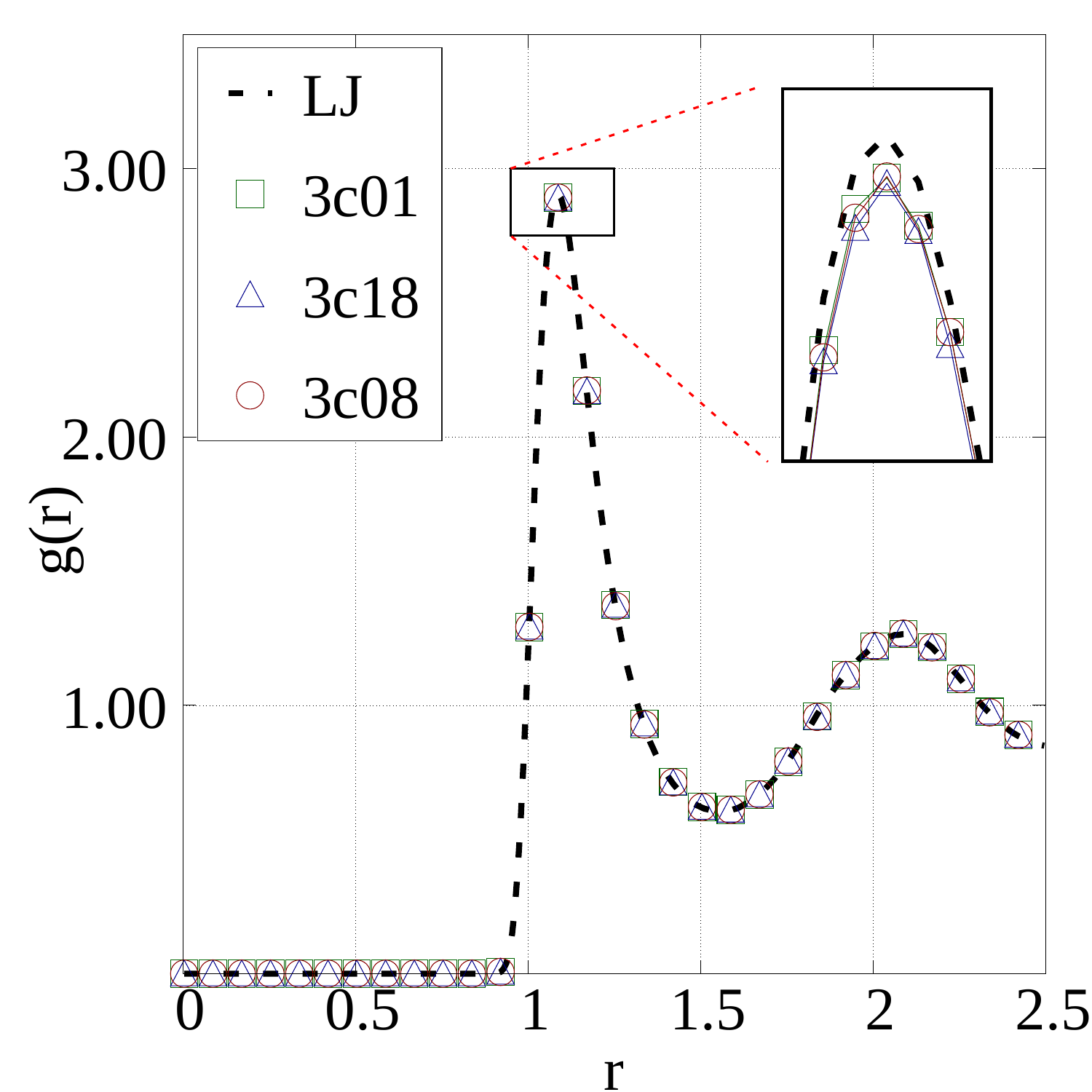}
				\\
				\multicolumn{2}{c}{(a) T=0.746, $\rho=0.817$}\\
				Pairwise & Product \\
				\includegraphics[width=0.49\textwidth]{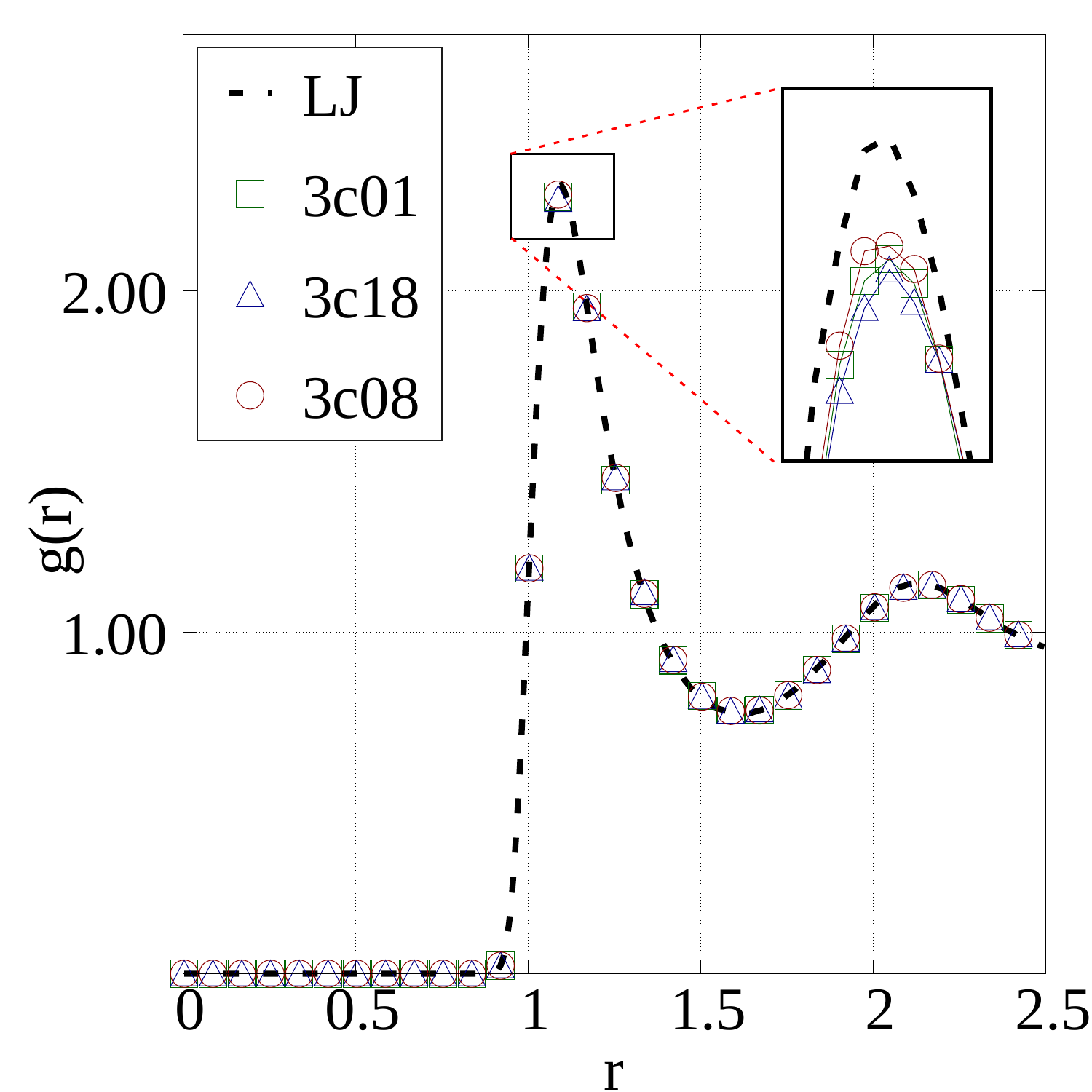}
				&
				\includegraphics[width=0.49\textwidth]{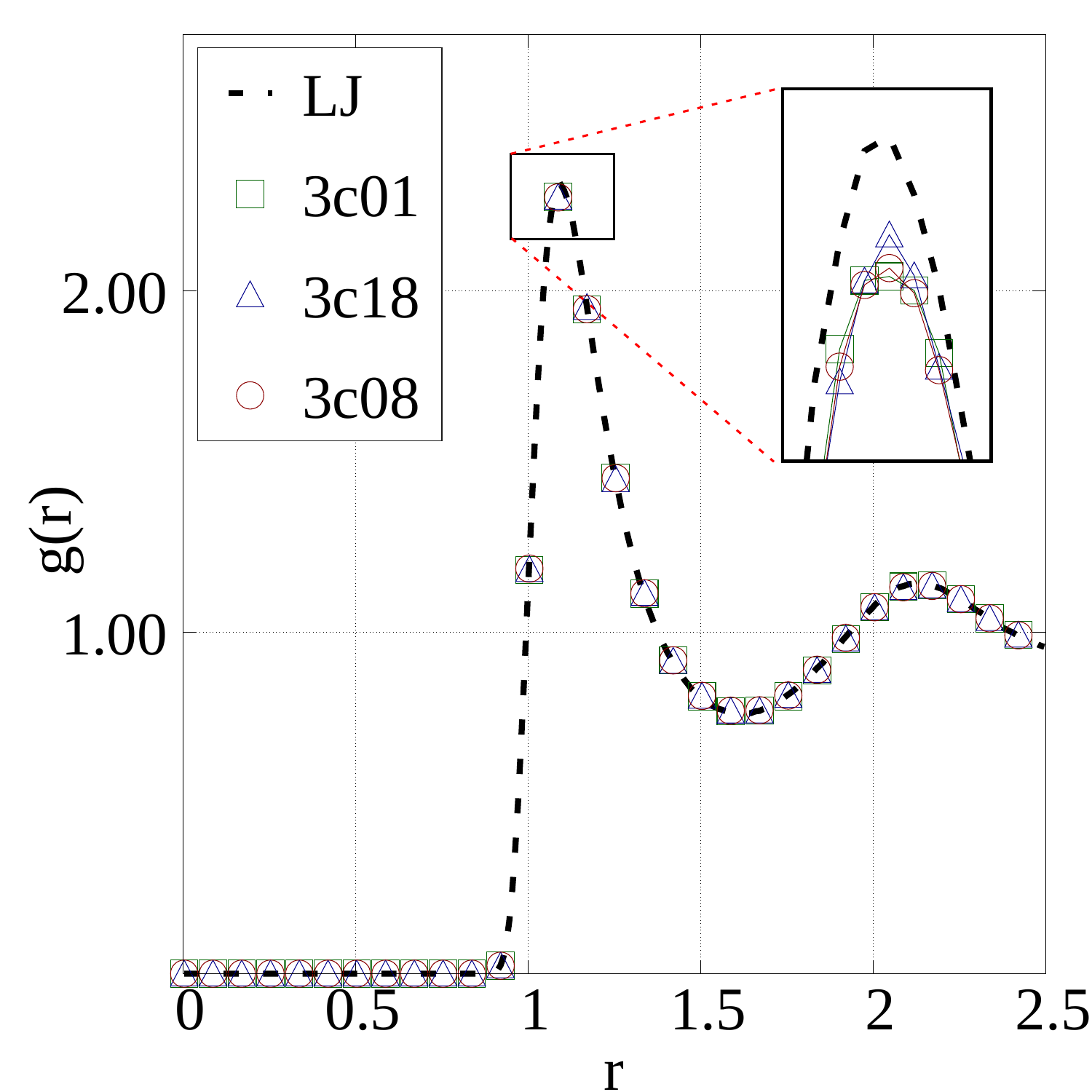}
				\\
				\multicolumn{2}{c}{(b) T=1.033, $\rho=0.65$}\\
			\end{tabular}
		\end{center}
		\caption{
			Radial distribution functions for (a) T=0.746, $\rho=0.817$ and (b) T=1.033, $\rho=0.65$ using the pairwise (left) and product (right) cutoff conditions. 
			The dashed line is the RDF obtained from a reference simulation run using only two-body Lennard-Jones interactions.
			A magnification of the RDF peak is given for each case.
		}
		\label{fig:rdfs}
	\end{figure}
	Additional to the ensemble averages, as part of our studies the radial distribution function was generated for all proposed traversal-cutoff combinations.
	These can provide a qualitative description of generated phase space using each traversal, where it is expected that all traversals allow for obtaining the same structural configuration.
	The radial distribution functions in \cref{fig:rdfs} are taken from the previously used thermodynamic states, T=0.746,$\rho=0.817$ and T=1.033,$\rho=0.65$.
	For completeness, the radial distribution function from the $\nu=0$ run is provided.
	In this sense, the resulting correlation functions can be compared between traversals and truncations, as well as to the Lennard-Jones scenario. 
	%It is assumed that the three-body forces have some effect on this correlation function, however this is not quantified in this work. 
	
	The main qualitative aspect of the results provided in \cref{fig:rdfs} is related to the obtained peak for each scenario, which is provided in higher detail.
	For both thermodynamic states, it can be seen that all traversals converge to the same structure irrespective of the used cutoff condition. 
	For the $T=0.746$ case, strong agreement between all traversals can be seen at the peak. 
	On the other side, for $T=1.033$ there are some minor differences at the obtained radial distribution function peaks.
	These hint to the presence of numerical differences introduced by traversing molecule triplets in a different order, stemming from each traversal.
	However, for both the pair and product condition the data points agglomerate around the same peak value.
	The additional molecule triplets introduced by the product condition do not seem to cause major structural differences in the obtained phase space.

	\subsection{Multiphase Simulations}\label{ss:multiphase}
	In this section we present results for the planar vapor-liquid interface. 
	As mentioned before, it is known from the literature that for simple liquids such as argon, simulation results for the computation of the surface tension obtained using only two-body interactions had a 19\% error with respect to experimental results~\cite{barker1993}.
	With the introduction of three-body interactions in the form of the Axilrod-Teller-Muto potential, this error is reduced to about 2.2\%. 
	In molecular dynamics simulations, a necessary step for the computation of the surface tensions is sampling a smooth density profile.
	With it, the interface configuration is to be found.
	For these reasons, numerical experiments were performed using the proposed traversals and cutoff conditions to generate the aforementioned density profiles.
	From them, the interfaces are analyzed in post-processing.
	
%	\begin{figure}[!htbp]
%		\centering
%		\includegraphics[width=0.9\textwidth]{figures/mp_rho/lj/lj-ref.pdf}
%		\caption{
%			Density profile at T=0.90 generated with only two-body Lennard-Jones interactions and force long-range corrections. 
%			The liquid and vapor interfaces are labeled accordingly and separated by interfacial regions. 
%			The interfaces are located at $z_0$ and have a width of $2d$, where $d=1.1607$ on the left and $d=1.2138$ on the right. 
%			The corresponding interfaces positions, $z_0$ are labeled, as well as the liquid and vapor densities, $\rho^l$ and $\rho^g$ respectively. 
%		}
%		\label{fig:ref-lj}
%	\end{figure}
	The simulations were setup following the literature, e.g.~\cite{walton1983,mecke1997}, by carrying out a 10,000 step equilibration phase of a homogeneous scenario with two-body and three-body interactions, the Lennard-Jones and Axilrod-Teller-Muto potentials respectively. 
	In all cases $T=0.90$, $\delta t=0.004$, and $r_c=2.5$.
	In the initial homogeneous equilibration, the box size is set to $\left(12.5\times12.5\times25\right)$ and a total of $N=3800$ particles is used.
	This is followed by an extension of the domain in the $z$-dimension, such that the length of the box in $z$ is $L_z=6a$, with $a$ the length of the other box sides and the dimensions are now $\left(12.5\times12.5\times75\right)$.
	In the inhomogeneous equilibration run, the initial velocities are chosen so that the system has no net momentum. 
	Additional long-range corrections to the two-body forces were implemented as given in \cite{janecek2006}.
	\added[id=JP]{No long-range corrections to the three-body forces were used.}
	The inhomogeneous phase space is equilibrated for 100,000 simulation steps, such that stable planar interfaces are formed.
	A density profile, $\rho\left(z\right)$, is sampled on every time step from this equilibration phase using a total of 600 bins over the $z$-dimension. 
	In post-processing, data fitting is used to determine the interface width $d$, position $z_0$, and the liquid and vapor densities, $\rho^l$ and $\rho^g$ respectively, using
	\begin{equation}\label{eq:tanh}
		\rho\left(z_i\right)= \frac{1}{2}\left(\rho^l+\rho^g\right)-\frac{1}{2}\left(\rho^l+\rho^g\right)\operatorname{tanh}\left(2\left(z_i-z_0\right)/d\right),
	\end{equation}
	where $\rho\left(z_i\right)$ are local density data points assigned to the center of each bin.
	
	\begin{table}[!htbp]
		\label{tab:rhoPars}
		\begin{center}
			\begin{tabular}{cc|ccc|ccc}
				%\hline
				&\multicolumn{7}{c}{T=0.90}\\
				%\hline 
				&&\multicolumn{3}{c|}{Pairwise Cutoff}&\multicolumn{3}{c}{Product Cutoff}\\
				Trav. &Par. &Left&Right&Mean&Left&Right&Mean\\
				%\hline
				\hline 
				\multirow{5}{*}{\coone}
				&$\rho^l$ & 0.7023 & 0.6995 & 0.7004 & 0.7008 & 0.6980 & 0.6994\\
				&$\rho^g$ & 0.0180 & 0.0183 & 0.0182 & 0.0161 & 0.0168 & 0.0165\\
				&$z_0$    & 20.5870& 54.2378&        &20.4065 &54.2099 &\\
				&$d$      & 1.3424 & 1.3961 & 1.3693 & 1.4322 & 1.3904 & 1.4113\\
				& & & & & & &\\
				\hline
				\multirow{5}{*}{\coneeight}
				&$\rho^l$ & 0.6979 & 0.6998 & 0.6989 & 0.7025 & 0.6968 & 0.6997\\
				&$\rho^g$ & 0.0176 & 0.0175 & 0.0176 & 0.0160 & 0.0161 & 0.0161\\
				&$z_0$    & 20.4505& 54.2160&        &20.5023 &54.3167 & \\
				&$d$      & 1.3628 & 1.3914 & 1.3771 & 1.3944 & 1.4038 & 1.3991\\
				& & & & & & & \\
				\hline
				\multirow{5}{*}{\czeroeight}
				&$\rho^l$ & 0.7031 & 0.6995 & 0.7013 & 0.7034 & 0.7004 & 0.7019\\
				&$\rho^g$ & 0.0163 & 0.0176 & 0.0170 & 0.0174 & 0.0180 & 0.0177\\
				&$z_0$    & 20.6001& 54.2787&        &20.6484 & 54.2526& \\
				&$d$      & 1.3861 & 1.3901 & 1.3881 & 1.4190 & 1.5354 &1.4772\\
				& & & & & & &\\
				&&\multicolumn{3}{c|}{\underline{Lennard-Jones($\nu=0$)}}&&&\\
				&$\rho^l$ & 0.7468 & 0.7484 & 0.7476 &&&\\
				&$\rho^g$ & 0.0122 & 0.0188 & 0.0155 &&&\\
				&$z_0$    & 21.5507& 53.3893&        &&&\\
				&$d$      & 1.1607 & 1.2138 & 1.1873 &&&\\
								
			\end{tabular}
			\caption{
				Parameters for the density profiles at $T=0.90$ for all traversal-cutoff combinations.
				Where appropriate, mean values for the left and right quantities are given.
				The Lennard-Jones, $\nu=0$, profile parameters are given as a reference and the left-side values are plotted in \cref{fig:density-profiles}. 
			}
		\end{center}
	\end{table}

\begin{figure}[!htbp]
		\centering
		\includegraphics[width=0.999\textwidth]{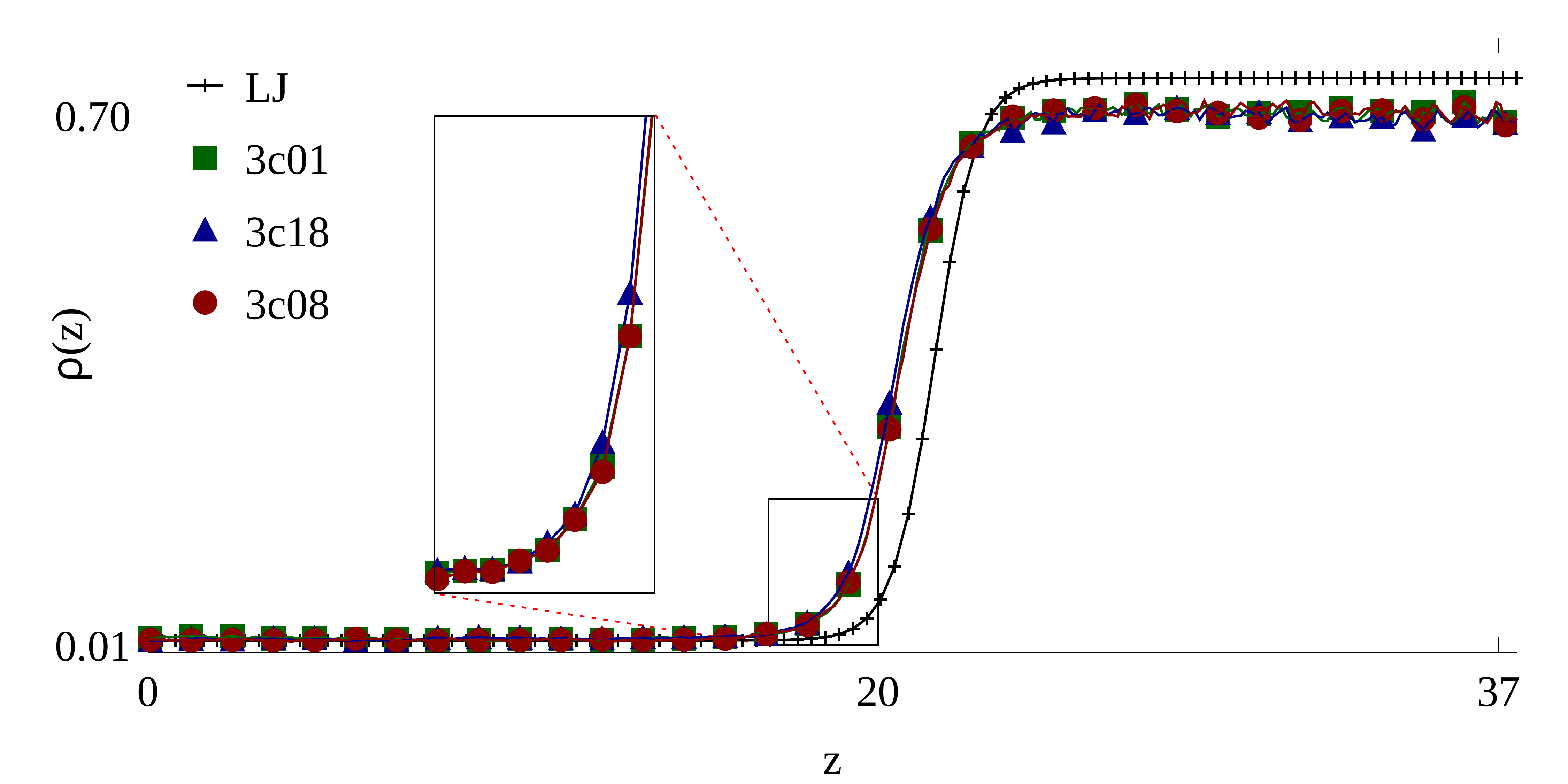}\\
		(a) Pairwise cutoff
		\includegraphics[width=0.999\textwidth]{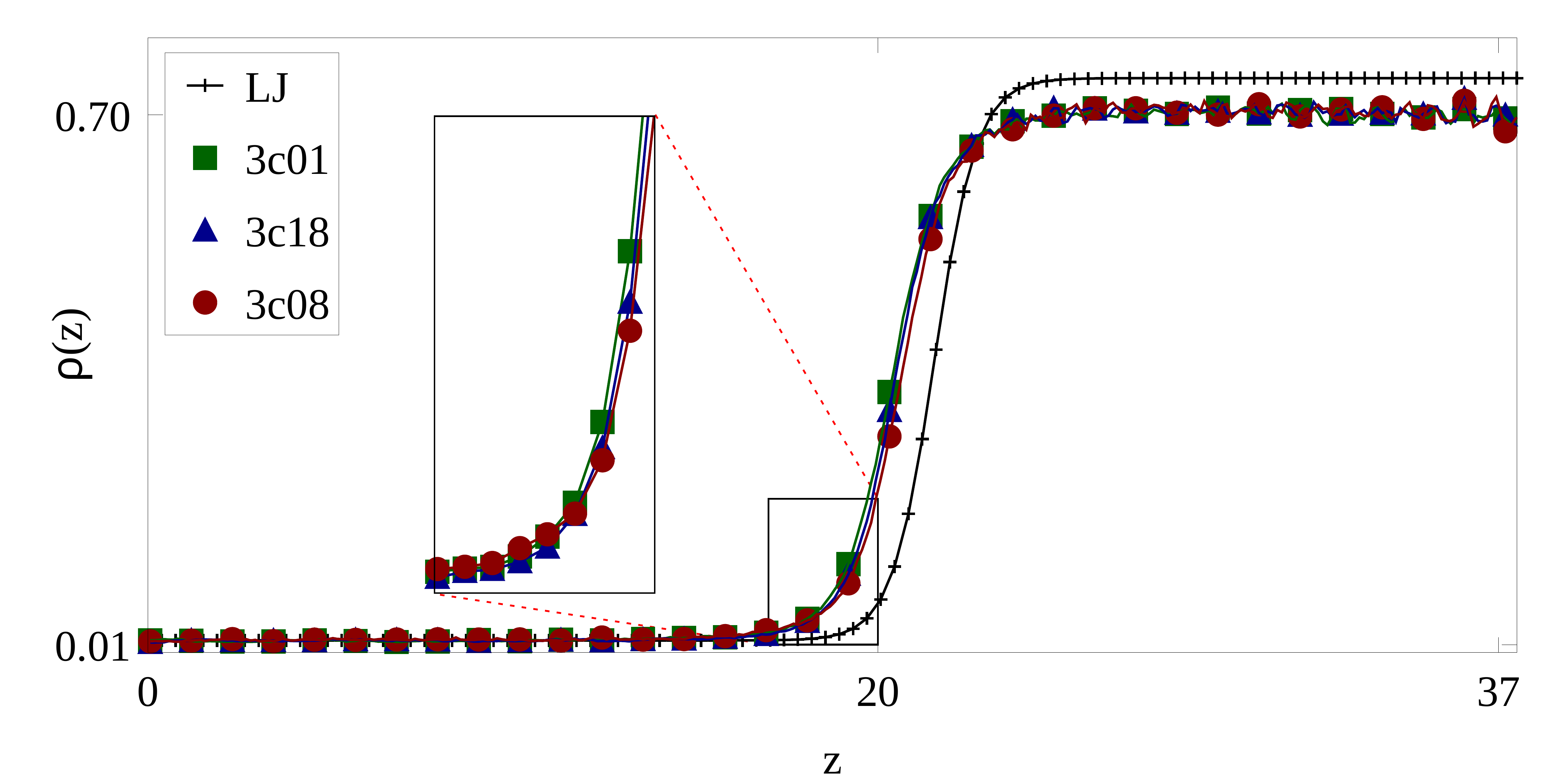}\\
		(b) Product cutoff
		\caption{
			The density profile for al traversals using the (a) pairwise and (b) product cutoff conditions are given. 
			The solid black line represents a reference Lennard-Jones profile interpolated via \eqref{eq:tanh}.
			For the three-body results, the interfacial region is magnified for an in-detail view of the data point overlap. 
		}
		\label{fig:density-profiles}
	\end{figure}

	\added[id=JP]{
	In the used scenarios, a liquid slab is surrounded by a vapor phase on both sides.
	The data sets for both sides were generated by fitting the left- and right-half data points to \eqref{eq:tanh}.
	The mean values are then reported when appropriate.
	In \cref{tab:rhoPars}, the interface parameters are reported.
	Finally, the left-side values were used to plot the density profiles in \cref{fig:density-profiles}.
	In order the illustrate the noticeable effects the Axilrod-Teller-Muto has on inhomogeneous scenarios, the parameters and density profile of a pure Lennard-Jones simulation are given in \cref{tab:rhoPars} and \cref{fig:density-profiles} respectively.
	This highlights the importance these three-body interactions have on more complex calculations such as the pressure tensor and the surface tension. 
	}
	
	\deleted[id=JP]{
	One critical aspect in inhomogeneous molecular dynamics simulations is related to the need of using long-range corrections to the forces in order to generate stables vapor-liquid interfaces, at short cutoff distances. 
	These are required because the truncation of the potential leads to instabilities between the phases.
	For a scenario with only two-body interactions, several approaches are available to allow for a short cutoff radius, e.g.~\cite{janecek2006,mecke1997}.
	This is not the case for three-body forces.
	This importance is exacerbated when using three-body forces, since the Axilrod-Teller-Muto decays slower than the Lennard-Jones potential.  
	Moreover, the velocities could be restarted to remove any net momentum generated during the equilibration. 
	Here, this has not been implemented, thus certain effects might be present.
	}
	\comment[id=JP]{Could this paragraph above be removed? Its more of a disclaimer? Or diluted into another one?}
	
	\added[id=JP]{
	The mean $\rho^l$ densities show good agreement between traversals, as shown in \cref{tab:rhoPars}. 
	For the vapor density, $\rho^g$, some disagreements can be observed between traversals and cutoff conditions.
	For the \czeroeight\, and product cutoff, this is to be expected since less triplets are traversed by design. 
	However, between \coneeight\, and \coone\, these differences are minor, particularly in the product condition.
	The results for the left and right interfaces positions, $z_0$, show very strong agreement in all cases, with some minor numerical differences in the decimal digits.
	In the same way, the interface widths show overall consistency.
	In the \czeroeight-product setup, the right interface width shows the largest differences with respect to the other configurations.
	Nevertheless, the mean $d$ value is still within the range of the other traversals. 
}

	\added[id=JP]{
	The parameter table of \cref{tab:rhoPars} is complemented by the density plots in \cref{fig:density-profiles}. 
	These are provided for the left middle of the domain, with a magnified view of the interfacial region. 
	The corresponding values of \cref{tab:rhoPars} were used, together with \eqref{eq:tanh}, to generate the plots.
	For all traversals, the data points overlap over the vapor and liquid phases. 
	In the case of the pairwise cutoff, at the interfacial region most data points overlap, with some differences seen in the magnified region. 
	Similarly, the profiles associated to the product condition show differences at the interfacial region.
	However, together with the parameters of \cref{tab:rhoPars}, it can be concluded that all the setups converge to the same interface geometry. 
	}
	
	\subsection{Traversal Performance Measurement}\label{ss:performance}
	The performance of the traversals is measured for a pure LJ fluid scenario.
	This was done for up to 64 cores in an AMD EPYC 7763 with  a single socket at 2.5GHz per core. 
	The strong scaling measurements are done using a 10 iteration run of a non-equilibrated phase space in an initial simple cubic configuration.
	
	The performance is measured in million molecule updates per second (MMUPS) using
	\begin{equation}\label{eq:mmups}
		\mathrm{MMUPS} = \frac{N\times \#{\mathrm{iterations}}}{T_{\mathrm{wall}}\cdot 10^6},
	\end{equation}
	where $T_{\mathrm{wall}}$ stands for the accumulated time it takes to compute three-body forces, given that it is the dominant task in the simulations. 
	%We take into account only the time it takes to compute three-body forces, because it is the dominant task in a simulation. 
	With this, the accumulated runtime during $10$ iterations is measured for every traversal, configured with each cutoff condition.
	A comparison is then drawn based on these measurements. 
	
	%	In order to compare systems of different sizes, the metric of million molecule updates per second, see~\cite{tchipevThesis}, is used
	%	\begin{equation}\label{eq:mmups}
		%		MMUPS = \frac{N\times \#{\mathrm{iterations}}}{T_{\mathrm{wall}}\cdot 10^6},
		%	\end{equation}
	%	where $N$ and $T_{\mathrm{wall}}$ give the total number of particles, and the measured walltime in seconds. 
	%	This metric is given in millions, $10^6$, for ease of notation. 
	%	In this way, the accumulated time for the computation of three-body forces across $10$ iterations is obtained and used in \eqref{eq:mmups} for the benchmarking of the traversals. 
	The simulations are set up with a total of $N=37,000$ molecules in a box domain with periodic boundary conditions on all sides and a box length $a=37.5$.
	Additionally, for all runs we use $r_c=2.5$ and $T=1.2$. 
	This gives a total of 3375 cells covering the complete domain, and the interactions are computed in parallel using the traversing algorithms and their respective coloring patterns. 
	OpenMP is configured using \texttt{spread} thread pinning, and all parallel loops use a \texttt{runtime} based scheduling.
	These settings were heuristically defined to be optimal in our experiments, via different trial runs with the mentioned simulation setup.  
	All measurements are averaged across five production runs to account for computational jitter. 
	
\begin{figure}[!htbp]
		\centering
		\begin{subfigure}[t]{0.9999\textwidth}
			\centering
			\begin{minipage}[b]{0.49\textwidth}
				\centering
				{\includegraphics[width=0.999\textwidth]{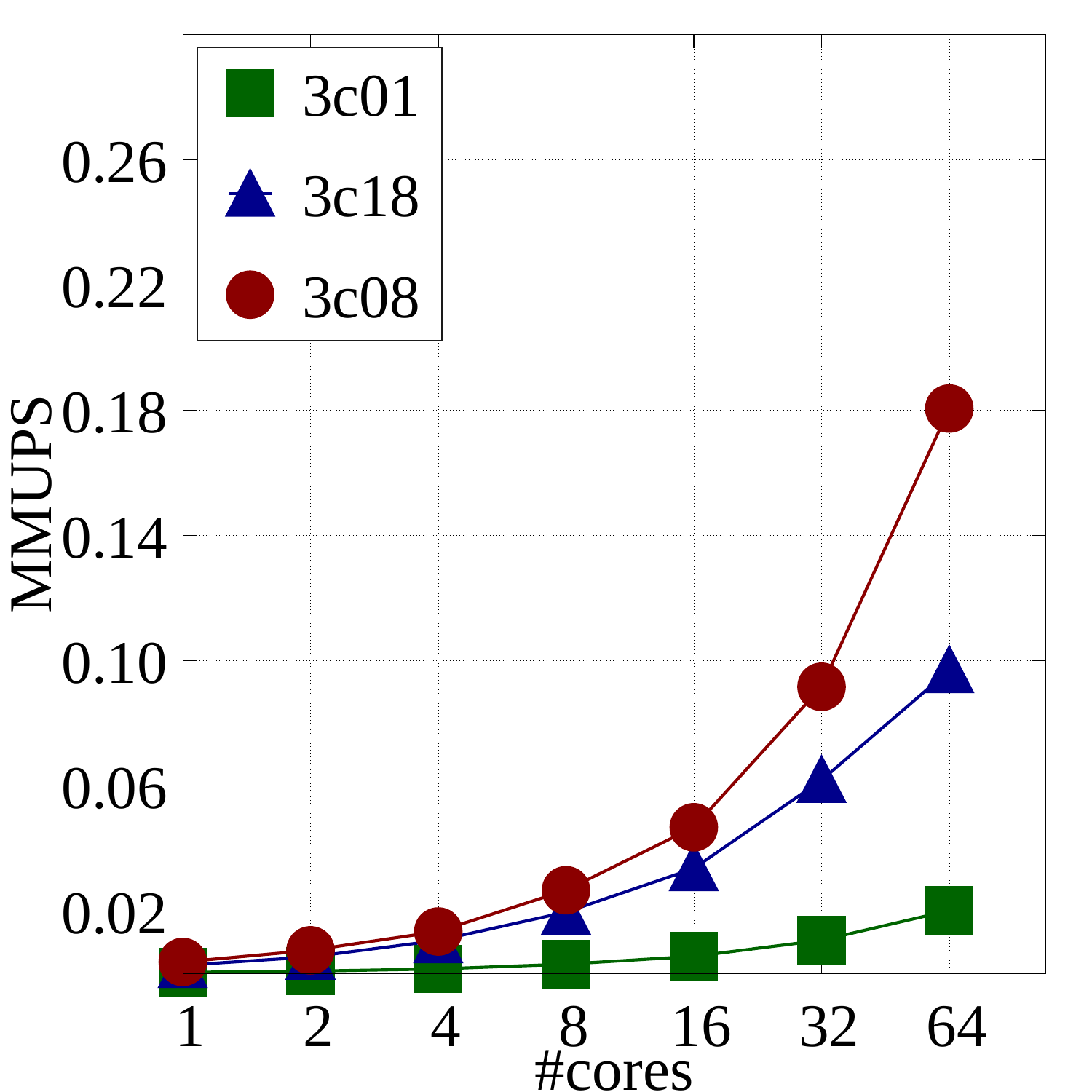}}
			\end{minipage}
			\hfill
			\begin{minipage}[b]{0.49\textwidth}
				\centering
				{\includegraphics[width=0.999\textwidth]{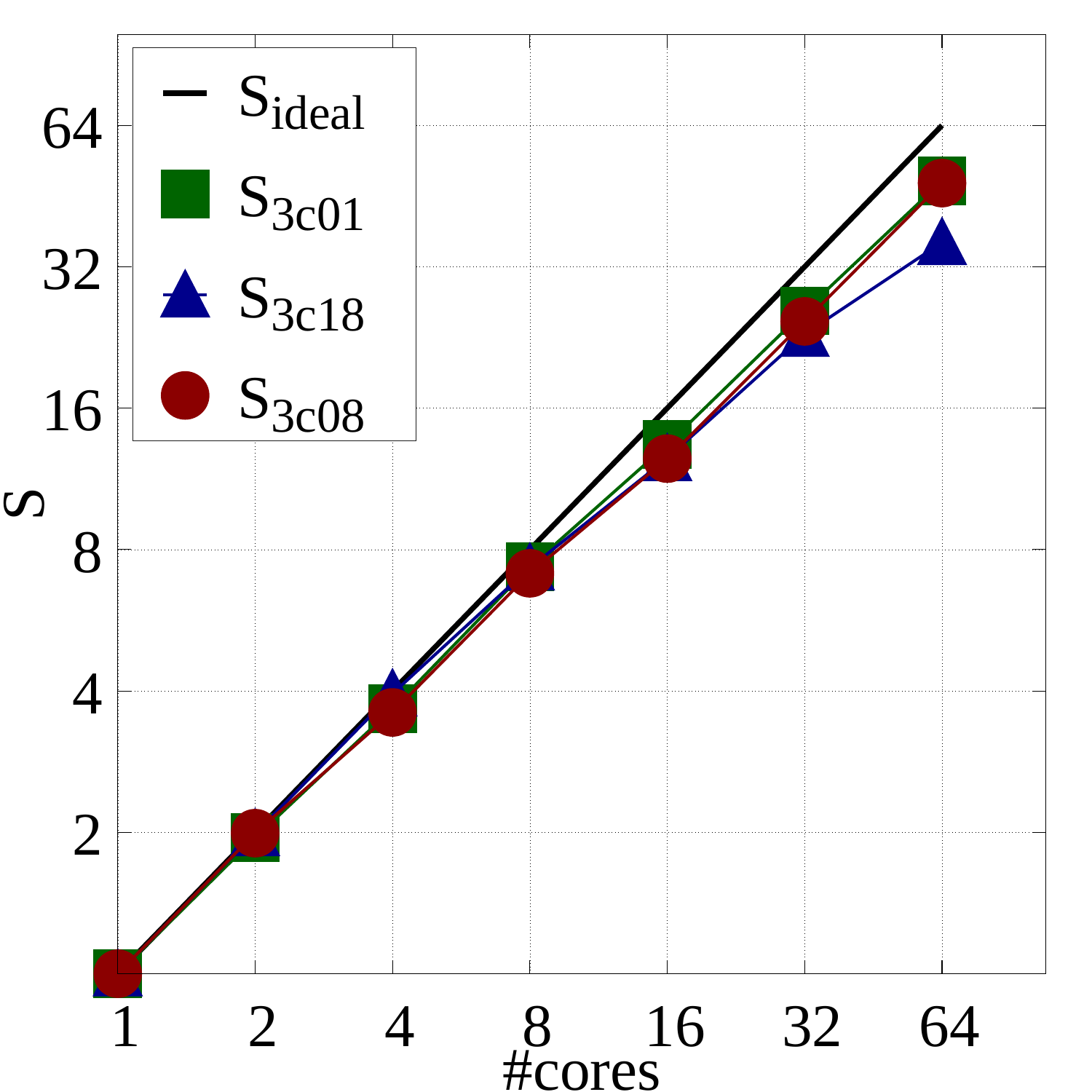}}
			\end{minipage}
			\caption{Product}
		\end{subfigure}	
		
		\begin{subfigure}[t]{0.9999\textwidth}
			\centering
			\begin{minipage}[b]{0.49\textwidth}
				\centering
				{\includegraphics[width=0.999\textwidth]{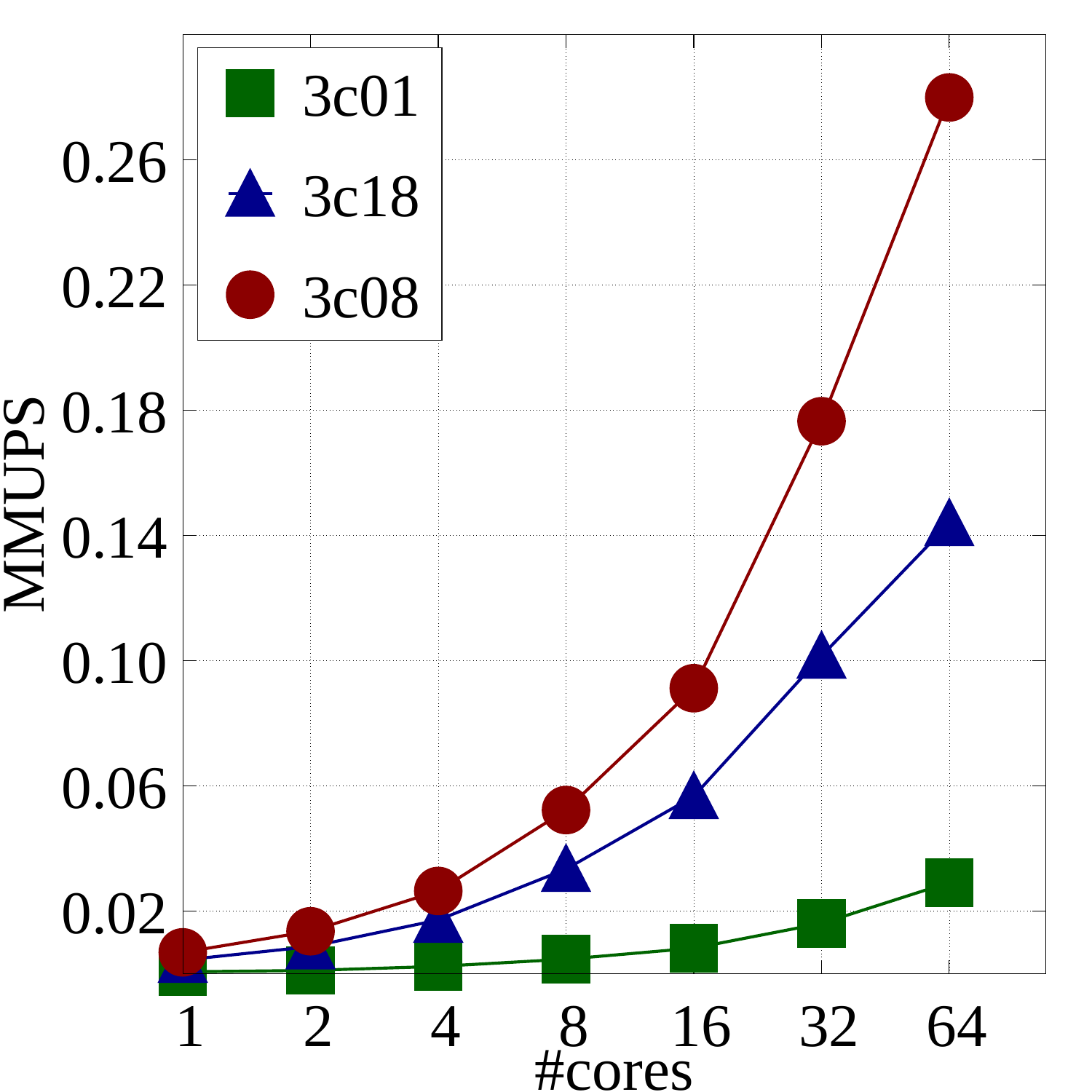}}
			\end{minipage}
			\hfill
			\begin{minipage}[b]{0.49\textwidth}
				\centering
				{\includegraphics[width=0.999\textwidth]{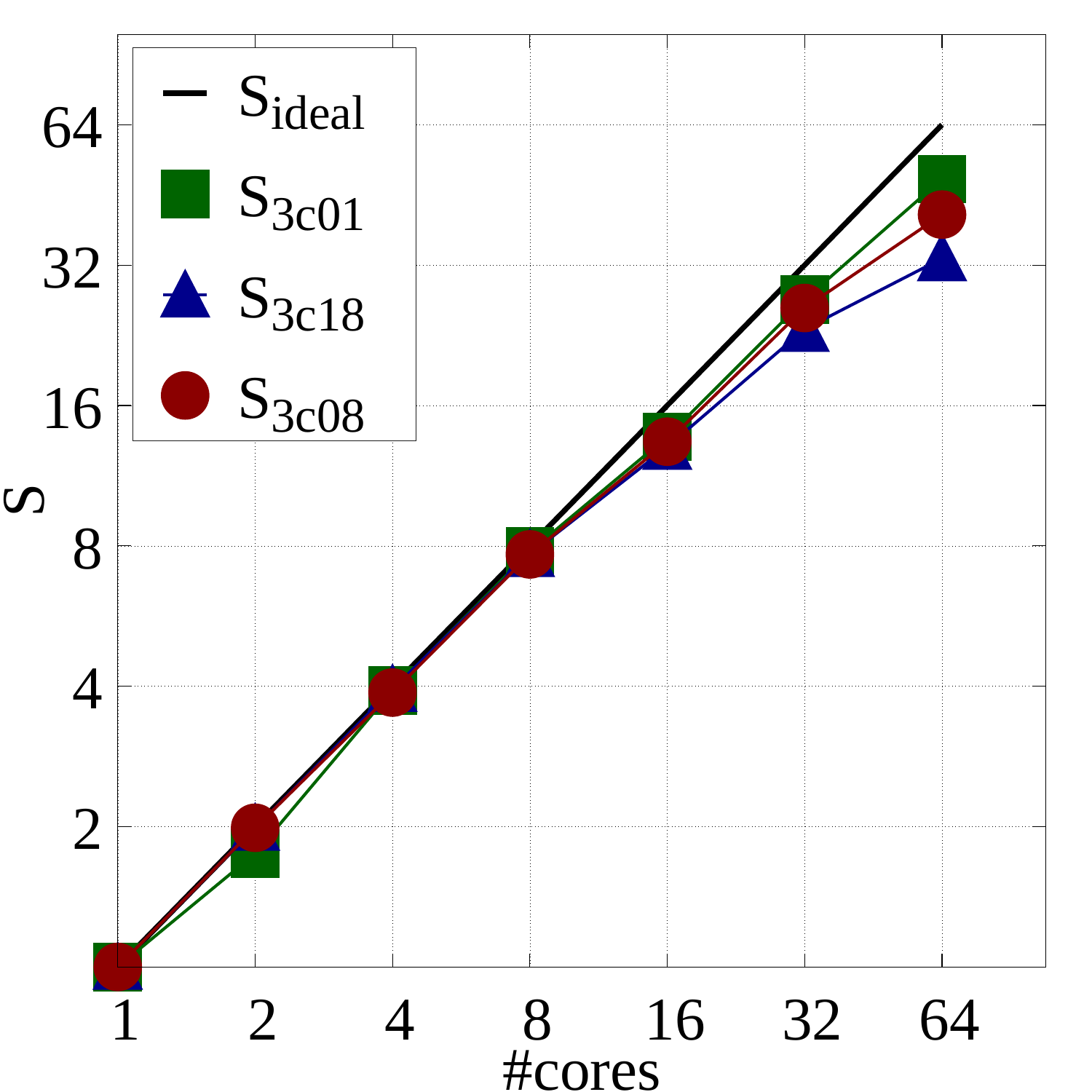}}
			\end{minipage}
			\caption{Pairwise}
		\end{subfigure}	
		
		%\vskip\baselineskip
		\begin{subfigure}[t]{\textwidth}
			\centering
			\begin{tabular}{c|rr|rr|rr}
				\#cores &  \multicolumn{2}{c}{\texttt{3c01}} & \multicolumn{2}{c}{\coneeight} & \multicolumn{2}{c}{\czeroeight} \\				
				\hline
				& Product      & Pairwise     & Product      & Pairwise       & Product      & Pairwise\\
				%				1  & 398.974544 &266.743256 & 63.305998 & 38.785404 & 45.984081 & 24.736635\\        
				%				2  & 204.676191 &134.860414 & 36.648681 & 19.895708 & 23.091777 & 12.449219\\    
				%				4  & 108.570980 &68.1516750 & 16.031418 & 9.9513260 & 12.994330 & 6.3961720\\
				%				8  & 54.4863990 &34.4060510 & 8.9799570 & 5.2533560 & 6.5149030 & 3.2392450\\
				%				16 & 30.2936920 &19.5521570 & 5.4191210 & 3.3280870 & 3.4702800 & 1.8603420\\
				%				32 & 15.4743420 &9.87377700 &3.4725650  & 2.1468450 & 1.9542040 & 1.0567990\\
				%				64 & 8.16827800 &5.43988200 &2.6170610  & 1.6894690 & 1.2754390 & 0.7854670\\
				%				1& 399.2161 & 266.6522 & 63.2557 & 38.8053 & 45.9712 & 24.7139 \\
				%				2&204.6723 & 135.9064 & 36.5799 & 19.9049 & 23.1037 & 12.4493 \\
				%				4&108.9211 & 68.2568 & 16.0498 & 9.8418 & 13.1356 & 6.3942 \\
				%				8&54.6184 & 34.5846 & 8.9728 & 5.2433 & 6.5544 & 3.2473 \\
				%				16&30.2892 & 19.5215 & 5.4350 & 3.3253 & 3.4935 & 1.8588 \\
				%				32&15.4765 & 9.9647 & 3.4686 & 2.1424 & 1.9530 & 1.0647 \\
				%				64&8.2144 & 5.4439 & 2.6855 & 1.6882 & 1.2785 & 0.7804
				1  &894.2426	&622.0282	    &137.7300	&85.0713	&98.8550	&54.2938\\
				2  &459.1410	&355.2080	    &69.2403	&42.8167	&49.6383	&27.3061\\
				4  &243.7996	&158.6939	    &34.8896	&21.5818	&27.4600	&14.0049\\
				8  &121.6006	&79.9605	    &18.8209	&11.0819	&13.8834	&7.0792\\
				16 &66.8196	&45.3539	    &11.0096	&6.5290	    &7.9129	    &4.0582\\
				32 &34.6914	&23.0465	    &5.9865	    &3.6462	    &4.0379	    &2.0970\\
				64 &18.3553	&12.7049	    &3.8193	    &2.5725	    &2.0499	    &1.3221
			\end{tabular}	
			\caption{Average accumulated walltime in seconds for 10 iterations}	
			\label{fig:runtimes}	
		\end{subfigure}
		\caption{
			Strong scaling results for the computation of the three-body forces using the cell triplet traversals.
			For each cutoff condition, product (a) and pairwise (b), the performance in MMUPS and the speedup for up to 64 cores are given.
			Additionally, the walltime in seconds (c) for the routine that computes three-body forces is given as the average between five production runs with 10 iterations each.
		}
		\label{fig:strong_scaling}
		%\end{adjustbox}
	\end{figure}

	%	
	%	\begin{figure}[!htbp]
		%	\centering
		%		\begin{subfigure}[t]{0.9999\textwidth}
			%			\centering
			%			\begin{minipage}[b]{0.49\textwidth}
				%				\centering
				%				{\includegraphics[width=1.0\textwidth]{figures/performance/3body_traversal_runtime.pdf}}\caption{}
				%			\end{minipage}
			%			\hfill
			%			\begin{minipage}[b]{0.49\textwidth}
				%				\centering
				%				{\includegraphics[width=0.999\textwidth]{figures/performance/3body_traversal_speedup.pdf}}\caption{}
				%			\end{minipage}
			%		\end{subfigure}	
		%
		%		\vskip\baselineskip
		%		\begin{subfigure}[t]{\textwidth}
			%		\centering
			%			\begin{tabular}{clll}
				%			\#cores & \texttt{c01} &\czeroeight & \coneeight\\
				%			\hline
				%			1  & 119.748406 & 24.7420670  &  38.881445\\
				%			2  & 97.8268660 & 15.0854390  &  19.563378\\
				%			4  & 48.9809060 & 6.37842700  &  9.8286810\\
				%			8  & 24.5170810 & 4.33037800  &  5.2663610\\
				%			16 & 12.8549470 & 2.46162600  & 3.3116140\\
				%			32 & 7.07074600 & 1.40835500  & 2.1130530 \\
				%			64 & 3.60181400 & 0.95159800  & 1.9275630
				%			\end{tabular}	
			%			\caption{}		
			%		\end{subfigure}
		%		\caption{
			%			Strong scaling results for the proposed traversals at node level. 
			%			Walltime in seconds for the computation of three-body forces using a scenario with $N=12,500$ and $\rho_N=0.8$ with $T=1.2$.
			%			The given times are given for a total of 10 iterations.
			%		}
		%		\label{fig:weak_scaling}
		%		%\end{adjustbox}
		%	\end{figure}
	The MMUPS, strong scalability, and average accumulated walltime in seconds are given in \cref{fig:strong_scaling}. 
	A higher runtime was measured for all cases using the product cutoff condition, as compared to the pairwise.
	Because this truncation admits a higher number of particle triplets compared to the pairwise condition, a higher workload at the same number cores was expected.
	For instance, for the \czeroeight\, case the runtime increases by almost a factor of two in some thread counts. 
	This also translates to better MMUPS for the pairwise condition in all cases, since on average, less forces need to be computed per particle.
	% Still, it is seen that in some cases the additional computations increase the runtime by almost a factor of two.
	% It could be speculated that this stems from the initial simple cubic structure, and the resulting particle distribution across the domain, which could result in a higher number of triplets across each cell.
	% However, this would require further testing. 
	% This fact, together with the minor differences observed in \cref{tab:1phaseRes} between the cutoff conditions, could offer a clearer picture of the trade-off between performance and accuracy. 
	%On the other hand, it has been shown in \cref{tab:1phaseRes} that the product cutoff condition does not result in large differences with respect to the pairwise.
	
	% Generally, a higher performance of the pairwise condition is also observed across all cases.
	% In terms of MMUPS, both the \coneeight\, and \czeroeight\, routines perform better than their product condition counterparts. 
	The highest MMUPS was measured for the traversal-truncation combination of \czeroeight-pairwise, as seen in \cref{fig:strong_scaling}. 
	At 64 cores, the \czeroeight\, routine had the lowest measured accumulated walltime during 10 iterations. 
	Nevertheless, in terms of strong scalability, we see that for the pairwise condition the \coone\, performed better than the other routines.
	This is because not using Newton's third law optimization makes the \coone\, traversal an embarrassingly parallel algorithm.
	Combined with any of the two truncation conditions, the scalability is close-to-ideal up to 64 cores.
	However, the results in \cref{fig:runtimes} show that the optimized \coneeight\, traversal allows for a roughly six-fold reduction of the runtime, which goes in line with the expected results as explained in \cref{sec:generalFrame}. 
	This improvement is even higher for \czeroeight\,, since fewer cell and particle triplets are traversed. 
	
	All of the above observations are also in line with the theoretical hitrate numbers given in \cref{tab:hitrate}.
	Going from \coone\, to \coneeight\, and then to \czeroeight\, the hitrate is increased by reducing the number of redundant distance checks, leading to a better performance.
	However, going from the pairwise cutoff condition to the product, the hitrate is increased by increasing the number of relevant triplets, leading to a worse performance.
	These results illustrate that the optimized traversals lead to a lower runtime, as well as to a more efficient use of the computing cores. 
	
	%	The MMUPS per traversal and truncation condition, together with the accumulated runtimes, are given in \cref{fig:strong_scaling}.
	%	The first evident observations, is that the product condition results in a significantly higher runtime than using the pairwise truncation.
	%	As mentioned, this scenario assumes a non-equilibrated phase space with a preexisting cubic structure.
	%	This allows for an evenly distributed number of particles per cell, so that no molecular clusters are present.
	%	This could magnify the effect of the product condition, since the additional triplets would be equally accounted for in every cell computation. 
	%	In comparison, the pairwise condition results in a reduction of runtime of up to half of the accumulated time, irrespective of the traversal. 	
		
	\section{Conclusion and Outlook}
	%Conclusion with summary
	We presented a general algorithmic perspective and cell traversals for the computation of three-body interactions in molecular dynamics simulations.
	Their design and implementation were described, together with their advantages and limitations. 
	These traversals have been benchmarked using Lennard-Jones fluids in homogeneous and inhomogeneous case studies.
	Good agreement between the results has been observed. \replaced[id=JP]{The homogeneous scenarios showed that the traversals, independently of which cutoff condition is used, can replicate results taken from the literature.
	The consistent radial distribution functions demonstrate that the effects of the traversals on the final phase space are minimal to non-existent.
	In the inhomogeneous case studies consistent interfaces were generated with all setups, thus allowing for the use of the proposed routines in more complicated numerical experiments such as the computation of the surface tension.
	}
	{although in the inhomogeneous scenarios some issues were observed in the vapor phase.
	These could be associated to the different number of triplets being considered by the different cutoff and traversal combinations.
	}

	The performance of all traversal-truncation combinations was measured at node-level for a meaningful number of particles.
	It was seen that, although the \coone\, traversal carries out calculations over repeated cell and particle triplets, its almost ideal strong scalability for up 64 cores could be exploited in massively parallel architectures. 
	Moreover, it was seen that the \czeroeight-pairwise simulation setup had the highest MMUPS and lowest runtime.
	This could be combined with distributed-memory parallelization to allow for three-body computations using a very large number of particles. 
	Furthermore, we could see that the hitrate is a good indicator of the performance of different three-body traversals.
	In addition to the performance, the observed strong agreement between the product and pairwise truncation conditions, particularly in the liquid state, indicates that the additional triplets included by the product condition might not be required in some applications.
	Thus, the workload can be further reduced.

	%Outlook
	Further node-level parallelization is to be achieved by the adequate vectorization of the force computation routines, based on the provided framework.
	Moreover, increased performance could be reached by properly balancing the threads to favor the more compute-intensive liquid state cells in the inhomogeneous case.
	Finally, these traversals can enable an in-depth study of phase behavior under the influence of three-body forces with a significant number of molecules.
	We aim to explore these aspects in future work. 
	\paragraph{Acknowledgments}
	\par
	{\small Computational resources have been provided by the project hpc.bw, funded by dtec.bw – Digitalization and Technology Research Center of the Bundeswehr. 
	dtec.bw is funded by the European Union – NextGenerationEU.
	The project has been supported by the BMBF project 3xa, 16ME0653, which has been financed by the European Union. }

	\printbibliography

\end{document}